\documentclass[sigconf]{acmart}

\usepackage[linesnumbered,ruled,vlined]{algorithm2e}
\usepackage{color}
\usepackage{multirow}
\usepackage{subfigure}
\usepackage{float}
\usepackage{textcomp}

\AtBeginDocument{%
  \providecommand\BibTeX{{%
    \normalfont B\kern-0.5em{\scshape i\kern-0.25em b}\kern-0.8em\TeX}}}

\setcopyright{iw3c2w3}
\begin{document}

\title{PARIMA: Viewport Adaptive 360-Degree Video Streaming}

\author{Lovish Chopra}
\authornote{Both authors contributed equally to this research.}
\affiliation{%
  \institution{Indian Institute of Technology}
   \city{Kharagpur}
   \country{India}
   \postcode{721302}
}
\email{lovishchopra98@gmail.com}

\author{Sarthak Chakraborty}
\authornotemark[1]
\affiliation{%
  \institution{Indian Institute of Technology}
   \city{Kharagpur}
   \country{India}
   \postcode{721302}
}
\email{sarthak.chakraborty@gmail.com}

\author{Abhijit Mondal}
\affiliation{%
  \institution{Indian Institute of Technology}
   \city{Kharagpur}
   \country{India}
   \postcode{721302}
}
\email{am@abhijitmondal.in}

\author{Sandip Chakraborty}
\affiliation{%
  \institution{Indian Institute of Technology Kharagpur}
   \city{Kharagpur}
   \country{India}
   \postcode{721302}
}
\email{sandipc@cse.iitkgp.ac.in}


\begin{abstract}
With increasing advancements in technologies for capturing {360\textdegree} videos, advances in streaming such videos have become a popular research topic. However, streaming {360\textdegree} videos require high bandwidth, thus escalating the need for developing optimized streaming algorithms. Researchers have proposed various methods to tackle the problem, considering the network bandwidth or attempt to predict future viewports in advance. However, most of the existing works either (1) do not consider video contents to predict user viewport, or (2) do not adapt to user preferences dynamically, or (3) require a lot of training data for new videos, thus making them potentially unfit for video streaming purposes. We develop \textit{PARIMA}, a fast and efficient online viewport prediction model that uses past viewports of users along with the trajectories of prime objects as a representative of video content to predict future viewports. We claim that the head movement of a user majorly depends upon the trajectories of the prime objects in the video. We employ a pyramid-based bitrate allocation scheme and perform a comprehensive evaluation of the performance of \textit{PARIMA}. In our evaluation, we show that \textit{PARIMA} outperforms state-of-the-art approaches, improving the Quality of Experience by over 30\% while maintaining a short response time. 


\end{abstract}


\copyrightyear{2021}
\acmYear{2021}
\acmConference[WWW '21]{Proceedings of the Web Conference 2021}{April 19--23, 2021}{Ljubljana, Slovenia}
\acmBooktitle{Proceedings of the Web Conference 2021 (WWW '21), April 19--23, 2021, Ljubljana, Slovenia}
\acmPrice{}
\acmDOI{10.1145/3442381.3450070}
\acmISBN{978-1-4503-8312-7/21/04}



\maketitle

 \section{Introduction}
{360\textdegree} videos have recently captured attention in the industry \cite{google-pixels, facebook-vr, alface2012interactive} as well as in academia \cite{corbillon2017360, fan2017fixation, park2019advancing, nguyen2018your, guan2019pano} due to their immersing and fascinating experience. Different video streaming platforms like Facebook and Youtube have introduced {360\textdegree} streaming as a part of their website and applications. However, one of the biggest disadvantages of {360\textdegree} video streaming is the large bandwidth requirement to provide a high-quality user experience. Since users have the flexibility to choose which part of the video they wish to see, the enriching experience comes with the cost of significant transfer of data, while only the part of the frame within the viewport of the video (\textit{Field of View} of the video player) can be seen by the users. Based upon general calculations, roughly {80\%} of the bandwidth is wasted during the streaming of a {360\textdegree} video, as a user rarely watches frames other than the viewports ~\cite{park2019advancing}. 

Due to large frame size, for a specific bandwidth, {360\textdegree} videos are streamed at a lower quality than a regular video. Thus, there is a need for optimisations over {360\textdegree} video streaming, which have emerged due to the large number of novel applications that they support \cite{fan2019survey}. The current standards for general video streaming over the web use \textit{HTTP Adaptive Streaming} (HAS) or \textit{Adaptive Bitrate Streaming} (ABR). During ABR, the streaming bitrates (quality) of the video frames are dynamically adapted based on the underlying network condition to ensure the best Quality of Experience (QoE) for the end users~\cite{kua2017survey}. Considering this, the optimisations during a {360\textdegree} video streaming can come from two fronts -- 
\begin{enumerate}
	\item[(a)] predicting the user viewports in advance so that the maximum network bandwidth can be utilised to stream the viewport part of the frames at the best possible quality (or bitrate), 
	\item[(b)] deciding bitrates for the viewports as well as for the non-viewports in the frames dynamically on-the-fly to maximize the end-users' QoE with the maximum utilisation of the available network bandwidth.
\end{enumerate} 

Many {360\textdegree} video streaming platforms \cite{google-pixels,facebook-vr} use tile-based streaming methods \cite{zare2016hevc, corbillon2017viewport, gaddam2016tiling}, where the frames of a video are spatially divided into $M \times N$ tiles and streamed as $MN$ chunks of tiles. Currently, most of the {360\textdegree} platforms like Youtube  and Facebook \cite{facebook-vr} use tiling-based methods for transferring the video frames. Recent researches~\cite{fan2017fixation, nguyen2018your,park2019advancing, clusterviewport} in the field of {360\textdegree} videos are focused on building models that can predict the viewport of a user to send only the part of the frame containing the viewport at higher quality and the rest of the frame at a lower quality, thus saving bandwidth. In other terms, at the same bitrate, a video can be viewed at a higher quality with viewport adaptive streaming since a higher proportion of the bitrate will be assigned to the viewport. However, many of these approaches for viewport-adaptive {360\textdegree} streaming do not utilise the exclusive video contents to predict the future viewports or adjust to user preferences during streaming or are not suitable from a streaming perspective~\cite{nguyen2018your, clusterviewport}. Consequently, the existing methods are mostly limited to specific types of videos and fail to satisfy all the primary goals of video streaming, such as maximum video quality at the viewports, smoothness in temporal scale as well as in the spatial scale (smoothness in the quality while moving from one tile to an adjacent tile), and minimum re-buffering latency. 

The user viewport depends on both (i) the content of the video and (ii) the personalised choice of the viewer, depicted by the past viewports. For example, in the case of a soccer match, the viewer may want to follow his/her favorite player or might be interested in the whole soccer field range, depending upon his/her personal choice. Similarly, for a concert video, the user might either focus only on the performer or might want to see the audience's reactions as well. Given such dynamic possibilities of having different viewports in the temporal scale for different viewers, viewport prediction in itself is a challenge. Consequently, such prediction would never result in a $100\%$ accuracy; therefore, during the {360\textdegree} video streaming, all the tiles of the frames need to be streamed, although the predicted viewport tiles may be streamed a higher quality than others. However, an abrupt quality change among the viewport tiles and the non-viewport tiles is also not desirable, as it would affect the video's spatial smoothness and thus, can affect the overall QoE. Finally, the prediction mechanism needs to be fast so that per-frame bitrates can be predicted in an online fashion during the videos' streaming. Given such multi-dimensional constraints, the overall optimisation of the {360\textdegree} video streaming is indeed a complex prediction and control problem. 

In this paper, we have developed a {360\textdegree} viewport-adaptive video streaming platform. The viewport prediction model on the client-side is online and adjusts to user preferences dynamically based on the video content. \textit{We claim that the viewport of a user depends upon the video contents along with the user's personal choice, based upon the movement trajectory of the prime objects in the video}. Our platform performs a one-time preprocessing of the video on the server-side to obtain the video's object trajectories. Our viewport prediction model, \textit{PARIMA}, which is an augmented combination of \textit{Passive Aggressive (PA) Regression} and \textit{Auto-Regressive Integrated Moving Average} (ARIMA) times series models, utilises the set of previously observed viewports and the object trajectories for the upcoming set of frames to predict the viewports for that set of frames, and incrementally learns the weights in an online fashion, thus adapting to user preferences dynamically. The augmented combination combines the benefits of the two individual models to create a model efficient for video streaming. Based on the predictions, the client allocates bitrates to each of the tiles using a pyramid-based allocation scheme, allocating a higher proportion of bitrate to the tiles corresponding to the predicted viewport and maintaining the QoE of the user. We do not model bitrate-adaptive streaming as a part of this research and assume the user bandwidth to be constant.

We evaluate our model on two publicly available data sets, one consisting of 5 videos with head movement data for 59 users, while the other consisting of 9 videos watched by 48 users, each video having a wide range of static and moving objects. We have made our code public\footnote{\href{https://github.com/sarthak-chakraborty/PARIMA}{https://github.com/sarthak-chakraborty/PARIMA}} for the research community. Using \textit{PARIMA}, we have achieved an average QoE improvement of around 35\% and 78\% over two baselines and an average improvement of 117\% in adaptivity over a non-adaptive bitrate allocation scheme. Our model is lightweight and exhibits a prediction latency of under 1 second for a chunk size of the same duration.

\begin{figure*}[t]
 \centering
   \subfigure[\textbf{Equirectangular image}]{\includegraphics[width=0.27\linewidth]{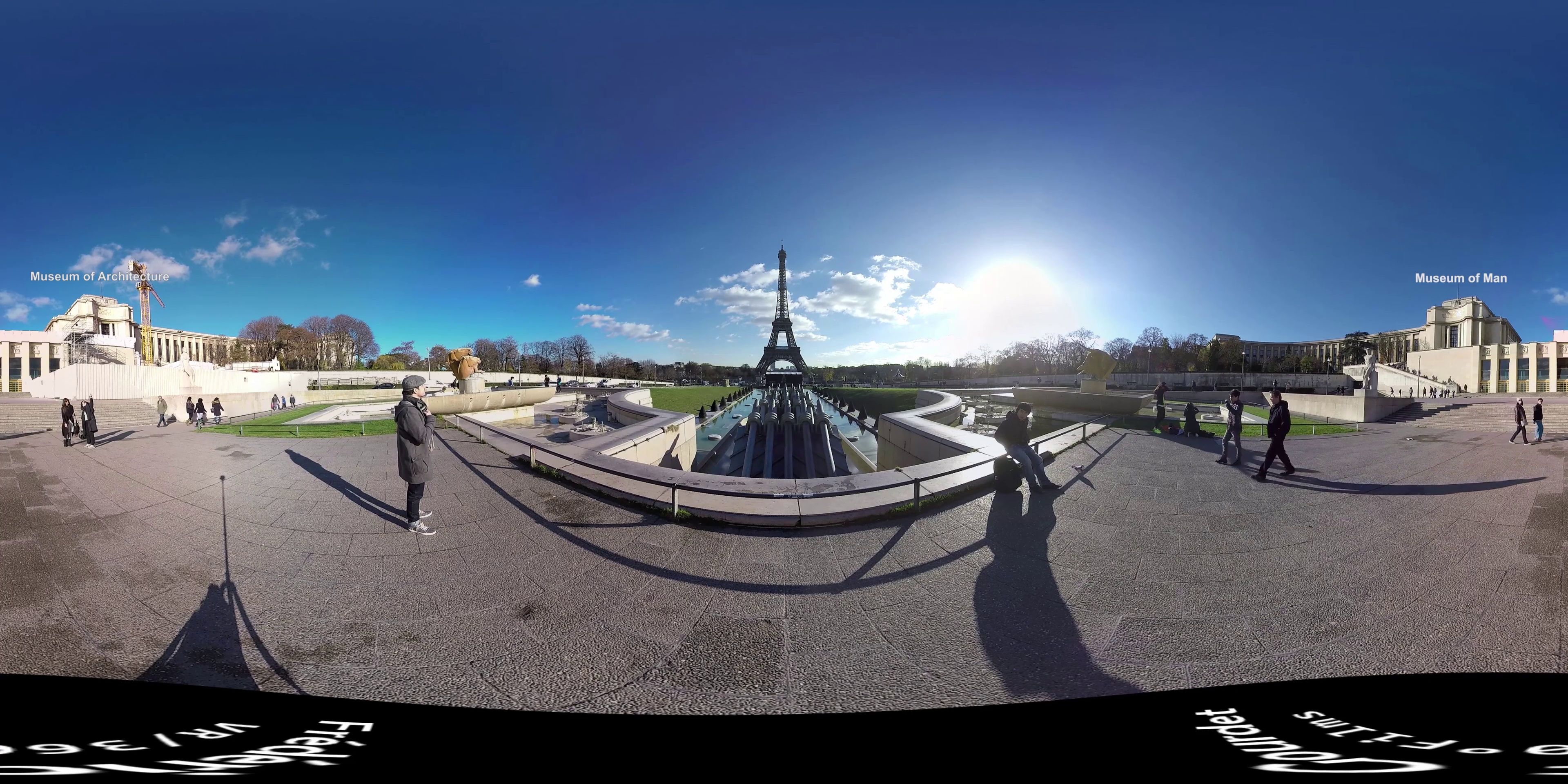} \label{eq}}
   \qquad
   \subfigure[\textbf{Cube map projection}]{\includegraphics[width =0.27\linewidth]{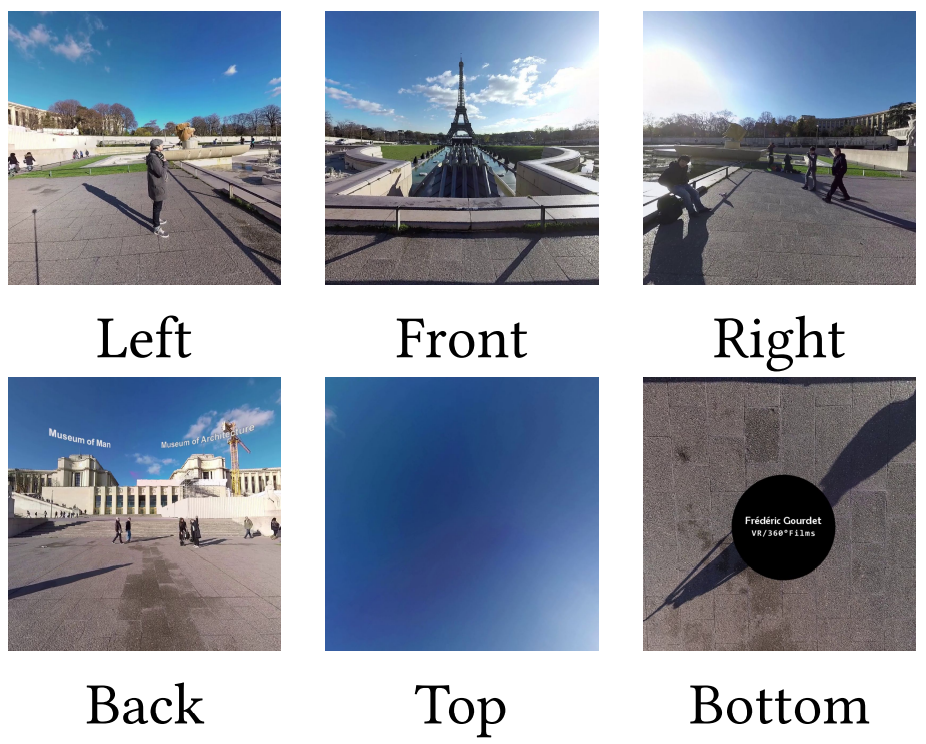} \label{cubemap}}
   \qquad
   \subfigure[\textbf{Stitched image with sample bounding boxes of objects detected}]{\includegraphics[width =0.27\linewidth]{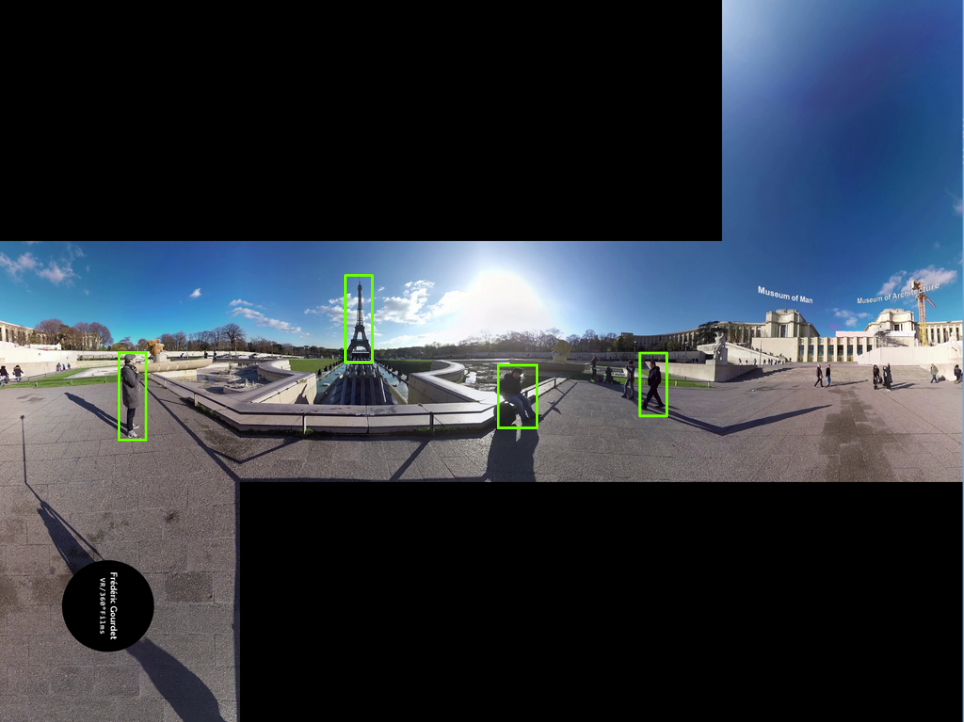} \label{stitchobj}}
\vspace{-5mm}
\caption{Figure shows the equirectangular projection, different faces of corresponding cube map projection and stitched image. Image is stitched such that the cube faces share common boundary. The bounding boxes in stitched image are examples for demonstration of object tracking. In reality, many more bounding boxes will be detected in the frame.}
\label{fig:stitching}
\end{figure*}

\section{Related Work}
In traditional HTTP-based adaptive streaming, a video is partitioned into temporal segments, and each segment is streamed with the desired quality to minimize the bandwidth requirement while maximizing QoE for the user, thus focusing on network congestion and available bandwidth. On the other hand, optimizations in {360\textdegree} video streaming involve an effort to reduce the streaming system's high bandwidth requirements by learning a model to predict the user viewports. The adaptations in {360\textdegree} video streaming involve viewport-adaptive and network-adaptive streaming techniques. Viewport-adaptive streaming aims to predict the future viewport of a user by learning user head movements to allocate specific parts of the frame with a higher bitrate. In contrast, network-adaptive streaming tends to model bandwidth fluctuations to utilize network bandwidth completely. Dynamic Adaptive Streaming over HTTP (DASH)~\cite{stockhammer2011dynamic} is a streaming standard that adaptively streams video based on the link bandwidth between server and client.

Any video streaming platform is typically a client-server system. For {360\textdegree} videos, due to the large frame size, each frame in the temporal chunks is further divided spatially into tiles~\cite{zare2016hevc, corbillon2017viewport, gaddam2016tiling}, and each of these chunks of tiles is stored at different bitrates on the server-side. Each frame may typically be divided into $64$--$100$ tiles, with each chunk of tiles being of usually $1$ to $4$ seconds~\cite{mpegdash}. At the back-end of the client-side, it requests the server for chunks of tiles at specific bitrates based upon the preferred quality and bandwidth available to the client. 
In adaptive bitrate streaming, H Mao \textit{et. al.}~\cite{mao2017neural} has proposed a reinforcement learning-based technique that learns the adaptive bitrate (ABR) algorithms adapting to a wide range of environment enhance user's quality of experience (QoE). It learns a control policy for bitrate adaptation from network throughput statistics and downloads the past few video chunks purely through experience. PARSEC~\cite{parsec} and SR360~\cite{chen2020sr360} uses a super-resolution on the client-side to stream video under constrained bandwidth, thus reducing bandwidth requirements and improving QoE for {360\textdegree} videos. 

Several studies have used viewport adaptive video streaming as a part of their research. Regression-based methodologies have been studied by~\cite{qian2018flare} and~\cite{bao2016shooting} where historical FoV trajectory is used. Works like~\cite{xie2018cls} and \cite{clusterviewport} cluster users periodically based on the head movement trajectory and assign new users to the existing clusters and predict the viewport. These, however, do not consider the use of video content and require an existing dataset of user viewports for any new video before predictions. Flocking based methodology is described in \cite{sun2020flocking}, which is applicable for a live {360\textdegree} video streaming where a large number of users are available concurrently. Recent studies like \textit{DRL360} \cite{zhang2019drl360} and \cite{chen2020sr360} have used deep reinforcement learning-based framework to predict viewport and optimize QoE objectives across a broad set of dynamic features. However, they don't consider video content while predicting viewports.

The existing literature has studied the saliency map concept to analyze the video contents~\cite{fan2017fixation, nguyen2018your, nguyen2013static}. A saliency map shows the properties of an image at the pixel-level, where a probability map over all the tiles is used, and the bitrate is decided based on this probability distribution. Fan \textit{et.al.} \cite{fan2017fixation} in his studies developed LSTM based model that learns the sensor-related features and image saliency map to predict viewer fixation in the future. PanoSalNet~\cite{nguyen2018your} learns the saliency map from user viewport data using DCNN and uses the LSTM network to predict the viewport. \textit{Mosaic}~\cite{park2019advancing} makes the use of a CNN + LSTM network to find a tile probability map using a saliency map and user head movement logs as inputs. However, learning a saliency map from head movements requires a lot of training data, making the model sensitive to extending to new videos. The use of LSTM models in the above works leads to a large number of parameters, and these systems do not update the parameters throughout streaming, leading to a lack of dynamic user adaptation, making them potentially unfit for adaptive video streaming.

This work addresses the issues of the previous studies and incorporates video contents, expressly object trajectories, and past viewports of the user to predict the next set of viewports. Though saliency maps had been used as a representative for video content, those have been generated only from the existing head movement data and not using the video content explicitly. Additionally for saliency maps, multiple areas of a frame can be predicted as salient (corresponding to multiple objects in the video), and hence, the tiles outside the viewport can be given high bitrate, resulting in possibly lower QoE.  Our model tends to these shortcomings, adapts to user preferences dynamically, is not heavily and statically parameterised and is easily extensible to new videos.

\section{Systems Overview}

\begin{figure*}[t]
 \centering
   \subfigure[\textbf{Bounding boxes from Figure \ref{stitchobj} re-projected to equirectangular frame and assigned initial object IDs}]{\includegraphics[width=0.27\linewidth]{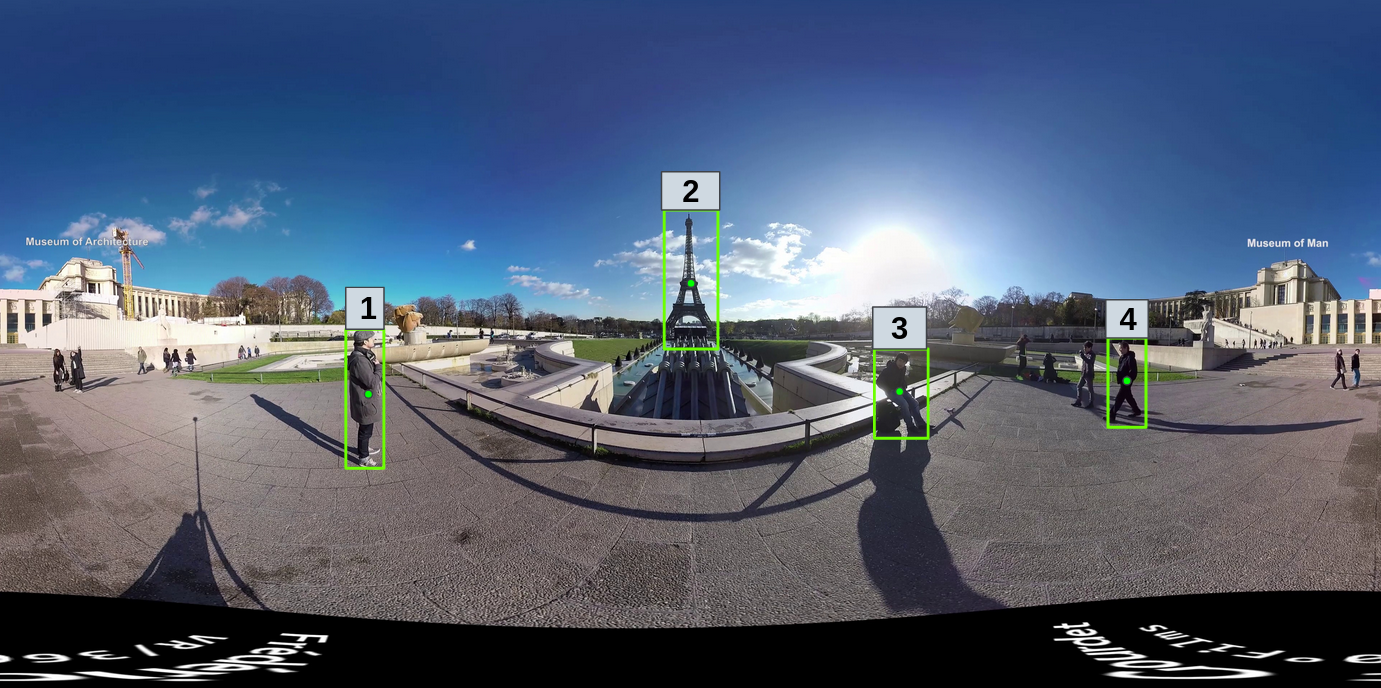} \label{eq_box}}
   \qquad
   \subfigure[\textbf{Indexing objects for next frame in spherical space}]{\includegraphics[width =0.27\linewidth]{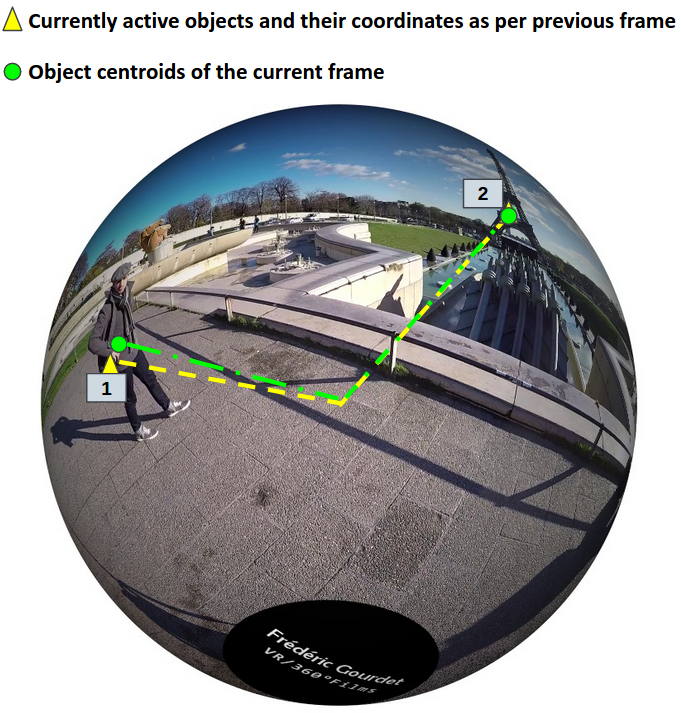} \label{spherical_track}}
   \qquad
   \subfigure[\textbf{Assigning object ids based upon the minimum spherical distance between active and next frame objects}]{\includegraphics[width =0.27\linewidth]{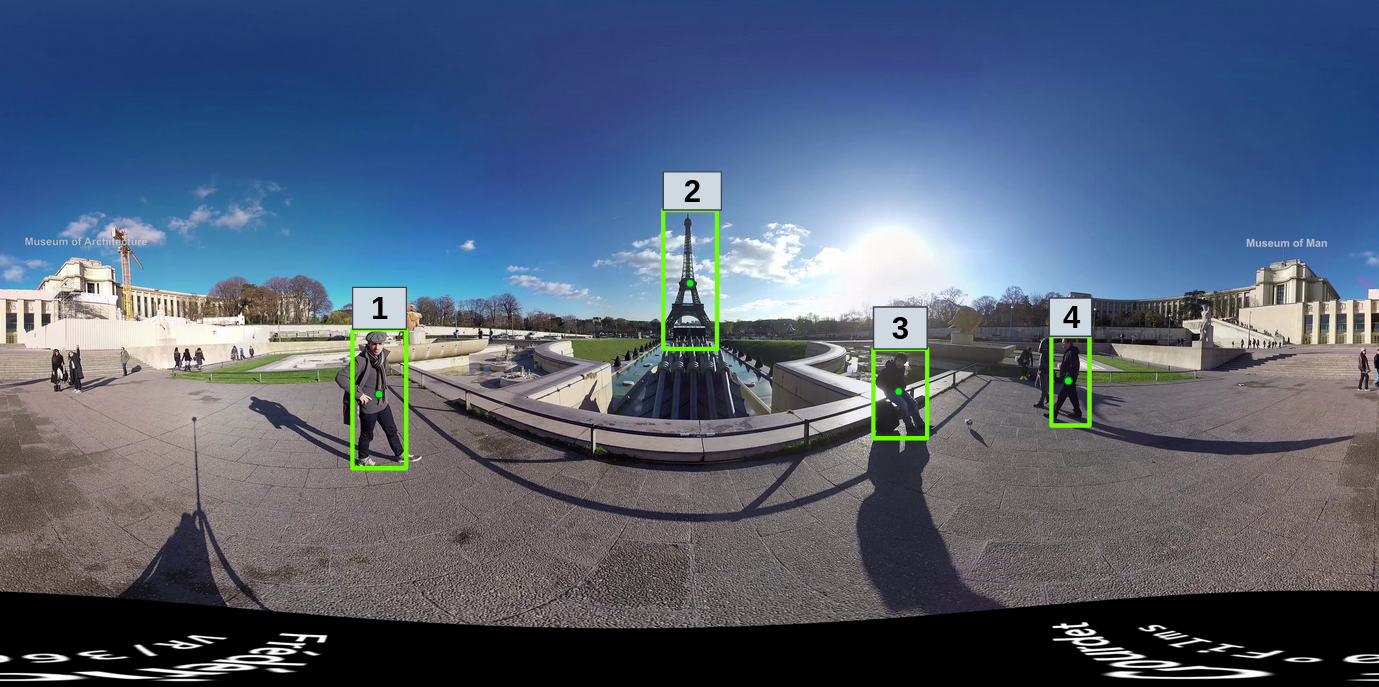} \label{new_obj_ids}}
\vspace{-4mm}
\caption{Spherical Object tracking for {360\textdegree} videos}
\label{fig:tracking}
\end{figure*}


Streaming {360\textdegree} videos involve a collection of tasks that need to be performed in order to provide the best user experience. Our methodology involves the use of video content to predict the future viewport of the user accurately. The task of finding the contents involves a detailed analysis of the trajectories of the objects present in the video and using them efficiently for viewport prediction. The user's current viewport is detected by capturing the head movement using the movement of the device playing the video.

\subsection{Video Preprocessing}\label{video_preprocessing}

Based upon our claim of user viewport depending upon the trajectories of prime objects in the video, we first run a one-time preprocessing of the {360\textdegree} video to obtain the object trajectories on the server-side. The object trajectory meta-data required for viewport prediction can be communicated to the client before streaming. Given the input video in the form of equirectangular frames \cite{equirectangular}, we index objects over the frames such that the same objects are assigned the same indices over multiple frames. The trajectory of an object is represented by the object index and a list of coordinates of the object over various frames. Existing object trajectory algorithms \cite{yilmaz2006object, han2004algorithm} cannot be used here since {360\textdegree} videos differ from general videos in various aspects, namely, (1) objects can wrap around an equirectangular frame to emerge from another side of the frame, and (2) the equirectangular frames are distorted and typically cannot be used for object detection using standard algorithms. Although there exist a few methods for object tracking over 360-degree videos, they either (1) work for single object tracking, or (2) run object tracking in equirectangular space. Consequently, the existing techniques fail in our case because we want more efficient multi-object trajectories for viewport prediction. Hence, we develop a robust methodology that can effectively track objects in {360\textdegree} videos, which we describe below. 

\subsubsection{Equirectangular to Cube-Map Conversion: }
{360\textdegree} frames in the equirectangular form are not ideal for image processing purposes because the frames are distorted in nature (Figure \ref{eq}). The distortion increases as we go near the poles. Hence, we convert the equirectangular frames to their cube map projection \cite{vrProjector}, which is the least distorted version of a {360\textdegree} frame as it projects the frame on six sides of a cube. The conversion is carried out by first converting the equirectangular frame to its corresponding spherical projection. In the second step, points in the spherical projection are then mapped to their corresponding faces in a cube-map. An example of the conversion from equirectangular to cube-map projection is shown in Figure \ref{eq} and \ref{cubemap}.

\vspace{-1.8mm}
\subsubsection{Frame Stitching and Object Detection: }
After converting the frames to their cube map projection with the distortion issue solved, we need to detect the objects present in each frame. However, in the conversion, since each pixel of an equirectangular frame is allocated a unique face of the cube, pixels of a single object might split and get mapped to different faces depending on its location on the sphere. Hence, if we run any regular object detection algorithm on each face of the cube separately, we might either not be able to detect the object or detect it as two different objects. To overcome this issue, we stitch the cube's different faces to form a single undistorted image (Figure \ref{stitchobj}). The stitching ensures that any object mapped to adjacent faces of the cube gets treated as an entire entity, ensuring that object detection and tracking are continuous. 
\par
We have used YOLOv3 \cite{redmon2018yolov3} algorithm on the stitched image of each frame to detect the objects and obtain their bounding box coordinates (Figure \ref{stitchobj}), which is effective in detecting objects due to the undistorted continuous nature of the image. We explicitly eliminated object classes because we can have multiple objects of the same class in a frame. The bounding box coordinates necessarily determines the current focus of the user among multiple available objects.
\par 
After the bounding box coordinates for each object is obtained, they are then reverse-translated back to their equirectangular projection. This is because the object tracking algorithm and the viewport prediction model run in the equirectangular space.

\subsubsection{Object Tracking: }
The final step in our pipeline for video preprocessing is to track the objects identified. Tracking objects essentially involves tagging them and assigning the same ID to `close' objects in successive frames. As discussed earlier, a major difference in {360\textdegree} videos is the flexibility of an object to wrap around the frame horizontally, which will be continuous in a spherical view but discontinuous in the equirectangular projection, in the case of which it should be assigned the same ID. Hence, we have devised a robust approximate spherical centroid object tracking algorithm that can index objects in {360\textdegree} videos efficiently. Here is a descriptive detail of the methodology used:

\begin{figure*}[t]
    \centering
        \includegraphics[width=\linewidth]{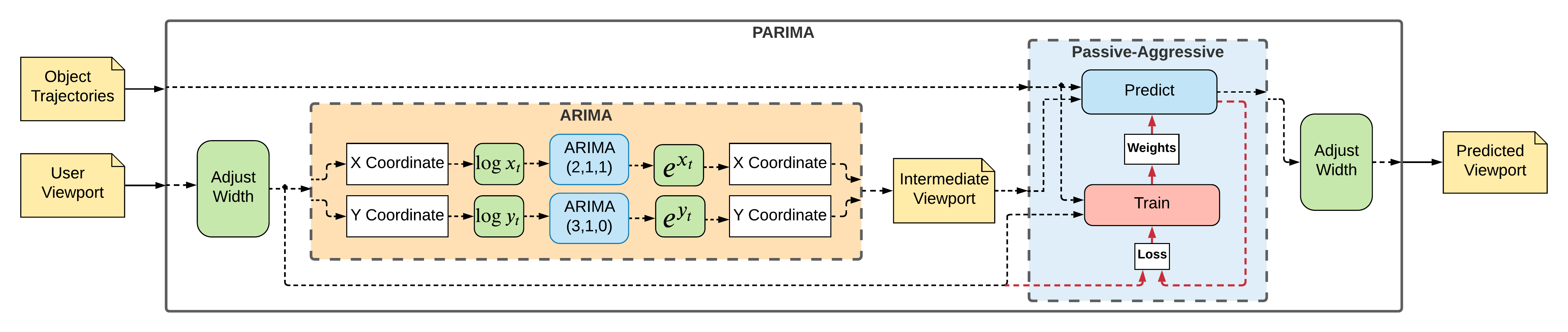}
        \caption{PARIMA Viewport Prediction Model}
        \label{fig:parima}
\end{figure*}

\begin{enumerate}
    \item Compute the centroid of the equirectangular bounding boxes of objects for all frames.
    \item For the initial frame, allocate each object a unique ID. These are the \textit{currently active} objects (Figure \ref{eq_box}).
    \item For the subsequent frames:
    \begin{enumerate}
        \item Project each centroid of the objects in the frame and currently active objects to spherical projection similar to Figure \ref{spherical_track} by casting the latitude and longitude to their corresponding spherical coordinates.
        \item For each pair of new frame centroids and active centroids, find the solid angle subtended by the spherical sector, which has the pair at diametrically opposite ends.
        \item For each object $O_i$ in the current frame, find the active object $O^{active}_j$, such that:
        \begin{equation*}
        \begin{split}
        & solidangle(O^{current}_i, O^{active}_j) <= solidangle(O^{current}_i, O) \text{ } \forall \text{ } O \in O^{active} \\ 
        & solidangle(O^{current}_i, O^{active}_j) <= solidangle(O', O^{active}_j) \text{ } \forall \text{ } O' \in O^{current}
        \end{split}
        \end{equation*}
        where $O^{current}$ is the set of objects detected in the current frame, $O^{active}$ is the set of active objects up to the previous frame. (Note that both ways need to be satisfied). Assign the $O^{current}_i$ with the same ID as the corresponding active object $O^{active}_j$ (Figure \ref{new_obj_ids}).
         
            \item \textit{Activate New Object}: If an object $O^{current}_i$ in the current frame has not been assigned any existing active object, it means that it has appeared for the first time in the video. In this case, we assign the object a new ID and set $O^{active} = O^{active} \cup \{O^{current}_i\}$
            
            \item \textit{Deactivate Old Objects}: If an active object $O^{active}_k$ is not assigned to any new object in the current frame, it means that either the object has disappeared or it went undetected. If the number of consecutive frames for which the object is not assigned any new object crosses a heuristic threshold of 30, we declare the object to have disappeared, and we set $O^{active} = O^{active} - \{O^{active}_k\}$.  However, if it reappears within 30 frames, it will be assigned the old object ID, the object coordinates for the missing frames are interpolated using the last available active and the new coordinates.
    \end{enumerate}
\end{enumerate}

Thus, we will obtain a set of centroid coordinates for each object index that was detected. The algorithm is robust because it uses the solid angle between two centroids in spherical projection and hence, takes care of the object wrapping issue discussed above, as well as interpolates object coordinates for missing frames. The window of 30 frames helps to overcome the inconsistencies of the YOLO algorithm where an object might go undetected for some frames in between.  
\par
Earlier, we have argued that stitching the different faces of the cube of a cubemap projection alleviates the problem of a distorted object either being detected as two separate object entities or not being detected at all. On the contrary, as shown in Figure \ref{stitchobj}, the `left' and the `back' faces of the cube were not stitched and an object transitioning among these two faces might still suffer the same fate. However, such minor inconsistencies are countered inherently in our object tracking algorithm, where a heuristic window of 30 frames along with spherical tracking technique helps to assign same IDs to the object for the missing frames. If the object is detected as two object entities (lower probability), one of them will disappear post-transition and will not be further considered by the model(Section \ref{viewportpred}). The use of previous viewport in the model further alleviates the problem, and hence we get a consistent representation of the video contents.

\subsection{Viewport Prediction}\label{viewportpred}

Predicting the upcoming viewport from the past viewports and the video metadata is a challenging task, especially when we want to model a dynamic system based on the videos' content. The learning task must be fast, accurate, online and should incrementally update the model weights to be able to adapt to user preferences quickly.
\par
An important and obvious requirement of the streaming model is to be able to predict multiple frames in the future at a single instance and stream them in the form of temporal chunks of specific $t$ seconds duration, rather than just predicting a single frame at a time, in order to maintain the QoE for the user. However, a larger chunk size can compromise the prediction accuracy, since user viewport tends to vary significantly within larger chunk duration, while these changes get reflected in the model only after the chunk is streamed. We evaluate the optimal chunk size duration in Section \ref{chunk_size_prediction}, based upon which, we use 1 second chunk size ( = $fps$ number of frames) for viewport prediction.  
\par
We have devised a model named \textit{PARIMA} (Figure \ref{fig:parima}), which is an augmentation of ARIMA \cite{arima} time-series model with Passive Aggressive Regression \cite{crammer2006online}, to predict user viewport effectively. The set of viewports for the next chunk of frames are predicted using the previous chunk's viewports and the object coordinates for the upcoming chunk of frames, obtained from Section \ref{video_preprocessing}.

\par
Time series models are often used in predicting head movements of users \cite{gaddam2016tiling}. Hence, it can also be used to predict the next viewport of a user while watching a {360\textdegree} video because the next viewport is essentially a temporal function of user head movement. 
The model takes as input the $x$ and $y$ coordinates (horizontal and vertical components respectively) for the viewports of the previous chunk of frames to predict the viewports for the next chunk of frames. Let $(x_f,y_f)$ be the viewport at frame $f$. However, if the viewport wraps around the equirectangular frame to another side of the frame, it would create a discontinuous time series of viewports. To model it as a continuous time series, $x_f$ needs to be width-adjusted, which is performed by the \textit{`Adjust Width'} component in Figure \ref{fig:parima} using the following transformation:
\vspace{-1mm}
\begin{equation}
    x_{f} : = 
\begin{cases}
  x_{f} + width & \text{if }[|x_{f} + width - x_{f-1}| < |x_{f} - x_{f-1}|] \\
  x_{f} - width & \text{if }[|x_{f} - width - x_{f-1}| < |x_{f} - x_{f-1}|] \\
  x_{f} & \text{otherwise}
\end{cases}
\label{eq:adjust_width}
\end{equation}

To overcome the case of a viewport component in the time series being negative and hamper further calculations, we shift the entire series to the right by $width$. It is to be noted that any viewport position $x_f$ is essentially same as $x_f + width$ for {360\textdegree} videos because of wrapping-around property.
\par
In order to remove inconsistencies from the data that can cause the entire viewport of the chunk from having identical values (thus generating a positive semi-definite auto-covariance matrix), we add \textit{random(0, 0.1)} to the viewport coordinates. Augmented Dickey-Fuller test \cite{dickey-fuller} showed that projecting the entire viewport data into logarithmic domain is necessary to maintain stationarity of the time series. The viewport coordinates $x_f$ and $y_f$ are then transformed to $log(x_f)$ and $log(y_f)$ respectively. These chunks of transformed x and y coordinates go into separate ARIMA models to obtain the future chunk of viewports in logarithmic domain, which are reverse-transformed to the viewport space by simple exponentiation (see Figure \ref{fig:parima}: \textit{`ARIMA'} block).

\par The above set of intermediate viewports obtained is fed as an input to the Passive-Aggressive Regression model\footnote{We have used \textit{creme} \cite{creme}, a Python library for online Machine Learning. We have modified the library code to accommodate our demands. One such change is fitting the model for $f$ frames at a single instance.}. \textit{General Model Definition}: Passive-Aggressive Regression \cite{crammer2006online} is an efficient online learning regression algorithm that computes the mapping  $f: \mathbb{R}^n \to \mathbb{R}$, $f(\mathbf{x};\theta) = \theta^T \mathbf{x}$ where, parameters $\theta,$ predictors $\mathbf{x} \in \mathbb{R}^n$. The algorithm uses the Hinge Loss Function, given by:
\begin{equation}
L(\theta, \epsilon) = max(0, |y-f(\mathbf{x_t};\theta)|-\epsilon)
\end{equation}
where $y$ is the actual value of the response variable.
The parameter \(\epsilon\) determines a tolerance for prediction errors. The weight update rule for PA Regression is:
\begin{equation}\label{eq:update_rule}
\theta^{t+1} = \theta^t + \alpha \frac{max(0,|y_t - \theta^T \mathbf{x_t}|-\epsilon)}{||\mathbf{x_t}||^2 + \frac{1}{2C}}sign(y_t - \theta^T \mathbf{x_t})\mathbf{x_t}
\end{equation}

We run a coupled Passive-Aggressive Regression model that predicts the $x$ and $y$ coordinates of the viewport for the next set of frames (see Figure \ref{fig:parima}: Passive-Aggressive Block). Along with the intermediate viewport, the model uses the object trajectories that were pre-calculated in Section \ref{video_preprocessing}. The equations for the predictions for each frame in the future chunk of viewports are given by:

\begin{equation}\label{eq:prediction}
\begin{split}
X_f' = \theta_{0x} + \theta_X . X_{f}^{ARIMA} + \sum_{i=1}^{N_{obj}} \theta_{ix} . O_{Xif} \\
Y_f' = \theta_{0y} + \theta_Y . Y_{f}^{ARIMA} + \sum_{i=1}^{N_{obj}} \theta_{iy} . O_{Yif}
\end{split}
\end{equation}
where $(X_f', Y_f')$ is the predicted viewport of the PARIMA model, $(X_{f}^{ARIMA}, Y_{f}^{ARIMA})$ is the intermediate viewport obtained for frame f using ARIMA model, $(O_{Xif}, O_{Yif})$ coordinates for the $i^{th}$ object for frame $f$ and $\theta$ values are the model parameters.
\par
Since the predicted viewport $(X_f', Y_f')$ was initially width-adjusted using equation \ref{eq:adjust_width} in order to take care of wrapping around of viewports, we apply a $\bmod \text{ } width$ on the x-coordinate of the predicted viewport to back-transform it within the equirectangular frame:

\par
When the next chunk of frames is rendered in the video streaming, the actual set of user viewports are obtained, and the weights of the object features of the Passive Aggressive Regression model are updated using the predicted and actual values of viewports as per the update rule in Equation \ref{eq:update_rule}. At the start of the whole process, we train the regression model with an initial $5*fps$ frames to prevent prediction from being 0 at the start. For these sets of frames, we can either download the whole equirectangular frame and assume the next frame's predicted view to be the same as the actual viewport of the previous frame. Figure \ref{fig:parima}, along with the above equations, shows the computation for each chunk of frames, with the same computation being repeated for all the chunks. It is to be noted that for each chunk, we create a new ARIMA model while the same Passive-Aggressive model is trained over various chunks and reused (Red arrows in Figure \ref{fig:parima} indicates that memory is preserved and the model gets trained via this path). The viewport prediction algorithm of \textit{PARIMA} is formulated in Algorithm \ref{alg:parima}.

\begin{algorithm}
\KwIn{Object Trajectories $O_f \text{ } \forall f$ frames, Streaming Viewport of the user $V_f \text{ } \forall f$ frames, PA model $M^{[1]}$}
\KwOut{Predicted Viewport for all frames}
Initial Training of $5 fps$ frames on model $M^{[1]}$\;
Initialise list of Predicted Viewport Chunks $PV$\;

\For{each chunk $c$}{
    Initialise list of Predicted Viewports for Chunk $c$: $PV_c$\;
    $F^{c-1} \gets$ list of frames in chunk $c-1$\;
    $F^c \gets$ list of frames in chunk $c$\;
    Initialise chunk size $cs = fps$\;
    $(X_{F^{c-1}}, Y_{F^{c-1}}) \gets$ horizontal and vertical components of $V_{F^{c-1}}$\;
    Adjust Width of $X_{F^{c-1}}$ according to Eq. \ref{eq:adjust_width}\;
    Make series $(X_{F^{c-1}}, Y_{F^{c-1}})$ stationary\;
    Initialise $M^{[2]}_x = ARIMA(2,1,1), M^{[2]}_y = ARIMA(3,1,0)$\;
    Train $M^{[2]}_x$ on inputs $X_{F^{c-1}}$ to get $X_{F^c}^{ARIMA}$, $M^{[2]}_y$ on inputs $Y_{F^{c-1}}$ to get $Y_{F^c}^{ARIMA}$\;
    
    \For{frame $F^c_f \in [F^c_1, F^c_{cs}]$}{
        $X'_{F^c_f} \gets M^{[1]}(O_{XF^c_f}, X_{F^c_f}^{ARIMA})$ according to Eq. \ref{eq:prediction}\;
        $Y'_{F^c_f} \gets M^{[1]}(O_{YF^c_f}, Y_{F^c_f}^{ARIMA})$ according to Eq. \ref{eq:prediction}\;
        Adjust Width of $X'_{F^c_f}$ by applying $\bmod width$ \;
        Append $(X'_{F^c_f}, Y'_{F^c_f})$ to $PV_c$\;
    }
    
    Append $PV_c$ to $PV$\;
    Train $M^{[1]}$ on inputs $O_{F^c_1} \to O_{F^c_c}$ and actual viewports $V_{F^c_1} \to V_{F^c_c}$\;
    
 }
\Return{$PV$}\;
\caption{GET\_PARIMA\_VIEWPORT($V, M$) \\ PARIMA based Viewport Prediction}
\label{alg:parima}
\end{algorithm}

\par

ARIMA time series model helps to maintain the locality information and gives smooth predictions, while  Passive-Aggressive Regression tries to update the model weights at any update step in such a way that the predicted value is as close to the actual value as possible, leading to better adaptation with fast iterations. The model update after every chunk ensures that it adapts to the user preferences dynamically. The coefficients of the object coordinates essentially represent the significance of that object, indicating user preferences.

\subsection{Bitrate Allocation}\label{bitrate_allocation_pyramid}
Once we obtain the predicted set of viewports for a chunk in the form of equirectangular coordinates, we map them to the appropriate tile number since we use tiling-based streaming for the system. As the next step, tiles in each frame need to be allocated bitrates based upon the available bandwidth. Bitrate allocation to tiles should be accomplished in a way such that the tiles corresponding to the viewport should get higher bitrate than the off-viewport tiles. It is also essential to note that for a particular chunk of frames, multiple tiles might be predicted as viewports since a user might span across multiple tiles within the chunk. Hence, all the tiles corresponding to the viewport need to be given a higher bitrate than the rest. The bitrate should reduce gradually as we move away from the viewport, to maintain an optimal experience for a user.

\begin{algorithm}
\KwIn{Predicted Tiles for chunk c: $T_{F^c}$, Preferred total bitrate: $B_p$}
\KwOut{Allocated Bitrates for chunk c: $B^c$}
Initialise Bitrate $B^c$\;
Initialise $\textit{weight}_{ij} = 1$ for each tile $(i,j)$\;
Initialise chunk size $cs = fps$\;
    \For{frame $F^c_f \in [F^c_1, F^c_{cs}]$}{
     $(i^\prime, j^\prime) \gets T_{F_f^c}$, predicted viewport tile\;
     $\textit{weight}_{i^\prime j^\prime} \gets \textit{weight}_{i^\prime j^\prime} + 1$\;
     \For{all $(i,j) \neq (i^\prime, j^\prime)$} {
        $d_{ij} \gets $ min. manhattan distance from $(i^\prime, j^\prime)$\;
        \If {$(i, j)$ is within Video Player FoV}{
            $\textit{weight}_{ij} \gets \textit{weight}_{ij} + 1 - \frac{d_{ij}}{2 * max(d_{ij})}$\;
            }
        \Else {
            $\textit{weight}_{ij} \gets \textit{weight}_{ij} + 1 - \frac{d_{ij}}{max(d_{ij})}$\;
            }
     }
    }
    $B^c_{ij} \gets \frac{\textit{weight}_{ij}}{\sum_{i,j} \textit{weight}_{i,j}} B_p \text{,    } \forall (i,j)$\;
\Return{$B_c$}\;
\caption{SELECT\_BITRATES($T_f, B_p$) \\ Assign Bitrates to Chunks}
\label{alg:bitrate}
\end{algorithm}

\begin{figure}[h]
    \centering
    \includegraphics[width=\linewidth]{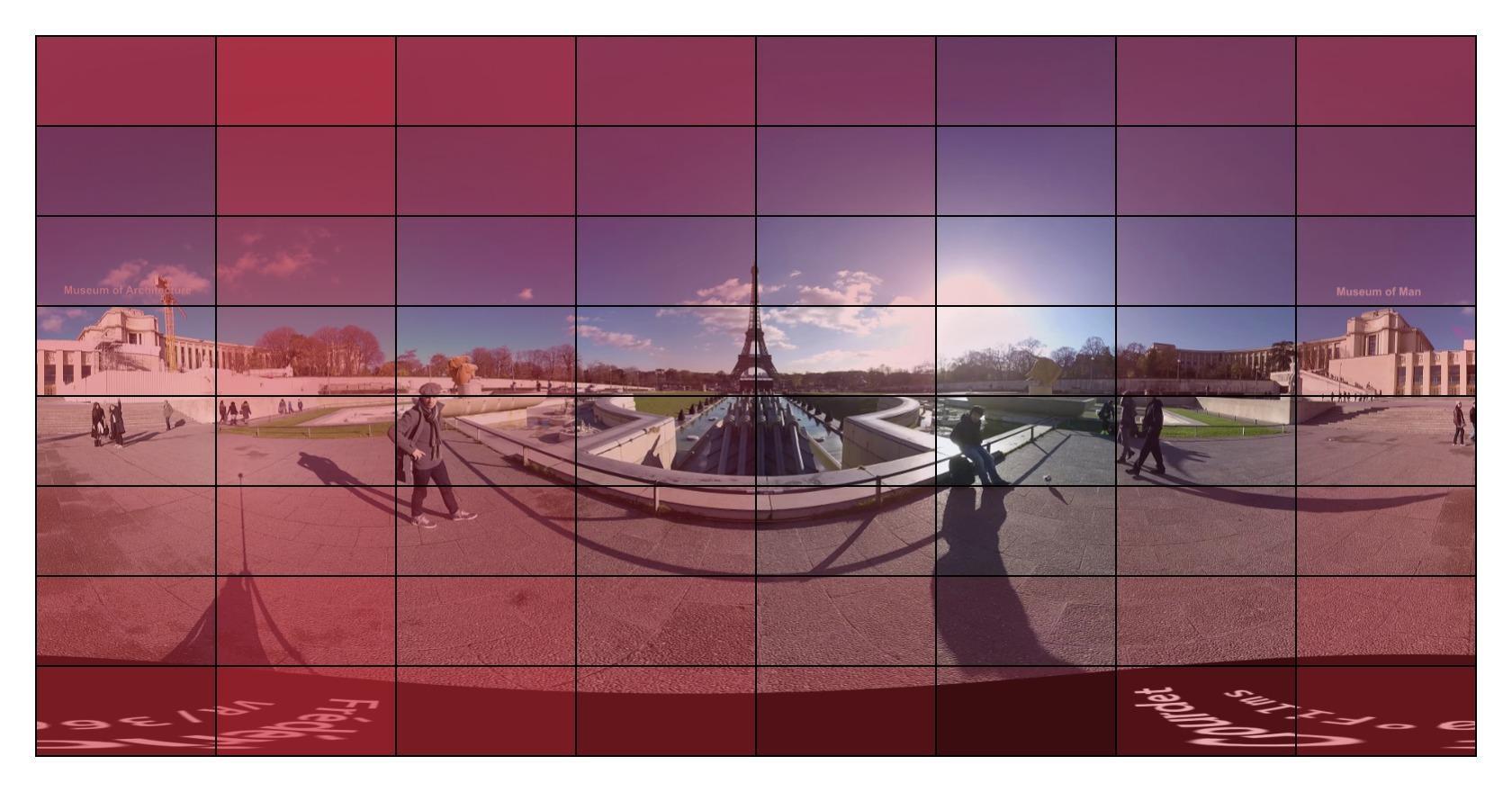}
    \caption{General Pyramid Model of Bitrate Allocation. The figure depicts allocation for a frame with $8 \times 8$ tiling, which has viewport at tile (6, 4). Lower opacity signifies higher proportion of bitrate to be allocated to the tile}
    \label{fig:Pyramid}
\end{figure}

\par
We have incorporated \textit{pyramid-based bitrate allocation scheme} \cite{gaddam2016tiling} in our model (Figure \ref{fig:Pyramid}). To assign a distribution of bitrates to the tiles $(i,j)$ in each chunk $c$, we use a weight function that would capture the proportion of total bitrate that should be given to a specific tile $(i,j)$. Whenever a tile $(i^\prime, j^\prime)$ is a candidate viewport, its weight is increased by a unit, and the weights of the other tiles are increased based on a pyramid approach, where the weight of the farthest tiles is increased by the least amount. Care is taken to allocate a higher bitrate for tiles within the Video Player FoV to maintain the Quality of Experience of the user. The distance of a tile from the viewport $(i^\prime, j^\prime)$ is measured as the minimum Manhattan distance because of the wrapping-around property of {360\textdegree} frames. Due to the wrapping around property, the maximum possible distance between two tiles is $(m + n)/2$, where the tiling is $m \times n$. The weight function is then normalized, such that the net weight across all tiles is one and then used to assign bitrates to each tile proportionally as formulated in Algorithm \ref{alg:bitrate}.

\section{Evaluation Testbed} \label{sec:testbed}
In this section, we give a brief description of the setup, datasets and the metrics used for the evaluation of our model. We have run the object trajectory algorithm on an Intel Xeon Gold 6152 processor with 88 cores (typical desktop hardware works as well), while the results for viewport prediction, bitrate allocation, and streaming client have been generated using simple desktop having Intel i5-4210U-quad-core processor and 8GB RAM. 

\par
\textbf{Datasets: } We use two popular datasets containing several 360-degree videos of different categories along with head tracking logs. 
The first dataset (ds1) \cite{corbillon2017360} includes five videos freely viewed by 59 users each with each video watched for 70 seconds. The second dataset (ds2) \cite{wu2017dataset} has nine popular videos watched by 48 users with an average view duration of 164 seconds. Each trace of the head tracking logs for both the datasets consists of the user head position in terms of unit quaternions $(w, x, y, z)$ along with the timestamp, which is converted to equirectangular viewport using the algorithm suggested by Nguyen et. al. \cite{nguyen2019saliency}\footnote{\url{https://github.com/phananh1010/PanoSaliency} (Access:\today)}. We have used the first 60 seconds of data for all the videos in our evaluation.

\par
\textbf{Baselines: }
The state-of-the-art baselines against which we have compared the performance of \textit{PARIMA} are described below:
\begin{itemize}
    \item \textit{PanoSalNet: }
PanoSalNet \cite{nguyen2018your} learns a panoramic saliency map for each {360\textdegree} frame by training a Deep ConvNet (DCNN) inspired architecture. The saliency maps are then passed along with user head movement data for the viewport prediction via an LSTM architecture. We have used the already available public code\footnote{\url{https://github.com/phananh1010/PanoSalNet} (Access: \today)} as our baseline.
    \item \textit{Cluster Viewport: }
This method \cite{clusterviewport} clusters users based upon their viewport history. It performs predictions for a new user by finding the cluster that the user belongs to and then use quaternion extrapolation to get the next chunk of viewports. For this approach, we have implemented the model with a prediction window of 1 second and have used \textit{pyquaternion}\cite{pyquaternion} library for quaternion-related calculations. From here on, we will refer to this model as \textit{Clust} for convenience.
    \item \textit{Non-Adaptive Bitrate Allocation (NABA) Model: }
To verify the adaptivity of our model, that is, whether \textit{PARIMA} can allocate bitrates intelligently to increase QoE, an important baseline to judge our model against is non-adaptive {360\textdegree} video streaming. Under this streaming model, there is no viewport-adaptation and hence, all tiles get an equal proportion of bitrate. In general, if $B$ is the preferred bitrate of streaming and the video is spatially divided into $M \times N$ tiles, then the bitrate allotted to each tile will be $B/(M \times N)$.
\end{itemize}
On one hand, while \textit{PanoSalNet} is a supervised learning strategy that uses saliency maps (and hence, indirectly video contents) to predict viewport, \textit{Clust} is a recent state-of-the-art algorithm, which uses unsupervised learning to cluster users and applies quaternion extrapolation for every chunk. Using the above mentioned viewport-adaptive streaming techniques having diverse methodologies yet being congruent to our study, we established that our choice of video content representation is more apt.
\par
\textbf{Metrics: }
\begin{itemize}
    \item \textit{Prediction Metrics: } For evaluating the accuracy of viewport prediction of our model, we have used Manhattan Tile Error as our metric. Tile error denotes the minimum Manhattan distance between the actual tile and the predicted tile, averaged over the video length.
    The Manhattan Error is reported as average over all frames and over all users for the video.
    \item \textit{QoE metrics: }User perceived quality is measured in a deterministic fashion using several QoE metrics that we define to empirically evaluate our model performance.
    \begin{enumerate}
        \item The first QoE metric $(Q_1)$ is the average bitrate consumed by the user in the actual viewport. In essence, it denotes the quality of the video perceived by the user. Let there be $\mathcal X \times \mathcal Y$ number of tiles in the video with a media player viewport dimension as $P_w \times P_h$. Mathematically, for chunk $c$, $Q_1^c$ is denoted as
        \begin{equation}
        Q_1^c = \frac{1}{n_c} \sum_{i=1}^{f_c} \big( \frac{ \sum_{P_w \times P_h} a^i_{x,y} B^c_{x,y}}{tiles(P_w)_n \times tiles(P_h)_n} \big)
        \end{equation}
        where, $f_c$ is the number of frames in the chunk $c$. $B^c_{x,y}$ is the allocated bitrate for $(x,y)^{th}$ tile in chunk $c$ and $a^i_{x,y}$ is an indicator variable which becomes 1 if tile $(x,y)$ is in the viewport of $i^{th}$ frame and 0 otherwise. $tiles(P_w)_n \times tiles(P_h)_n$ is the total number of tiles in viewed in video player.
        $n_c$ is a normalizing constant which is calculated by the number of distinct viewport tiles in chunk $c$.
        
        \item The second QoE metric $(Q_2)$ is a measure of the variation of the bitrate within the viewport for each frame. This metric ensures that we have minimum variation in bitrates of various tiles within the video player viewport. For chunk $c$, 
        \begin{equation}
        \begin{aligned}
        Q^c_2 = \frac{1}{n_c} \sum_{i=1}^{f_c} StdDev \{B^c_{x,y} :  \text{ } &x \in tiles(P_w),\\
        & y \in tiles(P_h)\}
        \end{aligned}
        \end{equation}
        
        \item The third QoE $(Q_3)$ captures the variation of bitrate among different frames for a chunk. We would like to minimize the amount of variation of quality from a viewport of frame $f_1$ to frame $f_2$. For chunk $c$,
        \begin{equation}
        \begin{aligned}
        Q^c_3 = \frac{1}{n_c} StdDev &\{ \frac{ \sum_{P_w \times P_h} a^i_{x,y} B^c_{x,y}}{tiles(P_w) \times tiles(P_h)}\\
        &:  a^i_{x,y}=1; \text{ }  \forall i \in f_c, \text{ } x \in \mathcal X, y \in \mathcal Y \}
        \end{aligned}
        \end{equation}
        
        \item The fourth QoE $(Q_4)$ captures the variation of viewport bitrate across successive chunks, which is essential to minimise for good Quality of Experience. For chunk $c$, 
        \begin{equation}
        Q^c_4 = |Q^c_1 - Q^{c-1}_1|
        \end{equation}
    \end{enumerate}
    
    For all the chunks C, the aggregate QoE is given by:
    \begin{equation}
    Q = \sum_{c=1}^C (Q^c_1 - Q^c_2 - Q^c_3) - \sum_{c=2}^C Q^c_4
    \end{equation}
    
    In our results, we report the QoE for videos averaged across all users.
\end{itemize}
\par
\textbf{Hyper parameters: }
\begin{itemize}
\item \textit{Tiling:} $8 \times 8$, amounting to total of 64 tiles. 
\item \textit{Chunk size:} 1 second, as discussed in Section \ref{viewportpred} and \ref{chunk_size_prediction}.
\item \textit{ARIMA Time Series Model Order (p,d,q):} $(2, 1, 1)$ for the x-coordinate and $(3, 1, 0)$ for the y-coordinate. This is achieved by hyper parameter tuning.
\item \textit{PA Regression Hyper parameters:} $C = 0.01$, $\epsilon = 0.001$. This is achieved by hyper parameter tuning.
\item \textit{Video Player Dimension:} 600 $\times$ 300
\item \textit{Preferred Video Bitrate by User:} Assumed to be constant at 8 Mbps (1080p) for the experiment. 
\end{itemize}

\section{Evaluation Results}
This section discusses the various experiments that we have performed to demonstrate the effectiveness of \textit{PARIMA}. We have compared our model against the baselines mentioned in Section \ref{sec:testbed}, as well as separately perform an individual assessment of the viewport prediction method. The results for the given datasets are generated using an emulator, which replicates the model to behave the same way as it would in a general {360\textdegree} setup. However, we also developed a basic video streaming system, where we have used a fast Kvazaar HEVC \cite{Kvazaar2016, hevc, zare2016hevc} encoder-decoder along with GPAC MP4Box \cite{gpac} for the creation of HTTP DASH segments, which ensures that the decoding time is low and is not a bottleneck in streaming. We use MP4Client \cite{gpac} for streaming client development.

\subsection{Optimal Chunk Size Prediction}\label{chunk_size_prediction}
This subsection discusses the choice of the optimal chunk size/prediction window for the model. Choosing an optimal chunk size essentially brings forth a trade-off between the streaming time and the prediction accuracy. An underlying trade-off also occurs where a client may need to store a considerable amount of video for a smooth experience, while prediction algorithms limit the amount of data to be stored in buffer~\cite{almquist2018prefetch}. A smaller chunk size facilitates better prediction accuracy and QoE for the user because of more frequent model updates. In contrast, a larger chunk size would suffer from a poor prediction model with low QoE. On the other hand, streaming time would have to be low for the chunks to ensure smooth streaming of video without buffering. We analysed \textit{PARIMA} for four different chunk duration: 0.5 seconds (chunk size: $fps/2$ frames), 1 second (chunk size: $fps$ frames), 1.5 seconds (chunk size: $3fps/2$ frames) and 2 seconds (chunk size: $2fps$ frames). 

\par
\begin{figure}[ht!]
        \centering
        \includegraphics[width=0.9\linewidth]{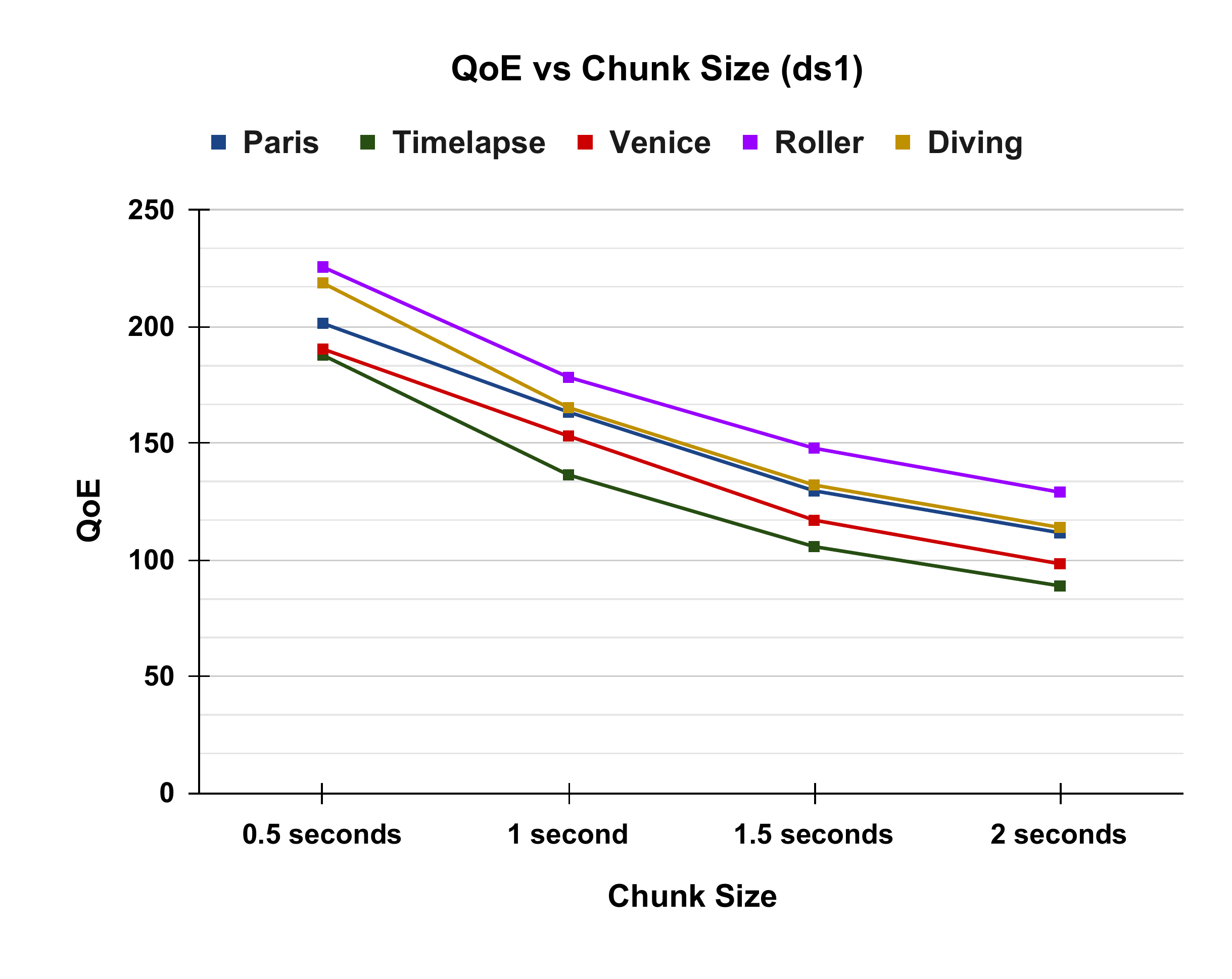}
        \caption{\textit{QoE of videos for varying chunk sizes}}
        \label{fig:chunksize}
\end{figure}

 To justify our claim of shorter chunks producing better QoE, we run the model on the five videos in dataset ds1 for the various chunk sizes and report the QoE for each video, averaged across all chunks for all users. As expected based upon the above argument, Figure \ref{fig:chunksize} shows a strict decrease in the average QoE for all the videos as we increase the chunk size. Similar results were obtained for dataset ds2. Hence, the smaller chunk size is always preferable over larger ones in terms of QoE.

\begin{figure}[ht!]
 \centering
   \subfigure[]{\includegraphics[width=0.9\linewidth]{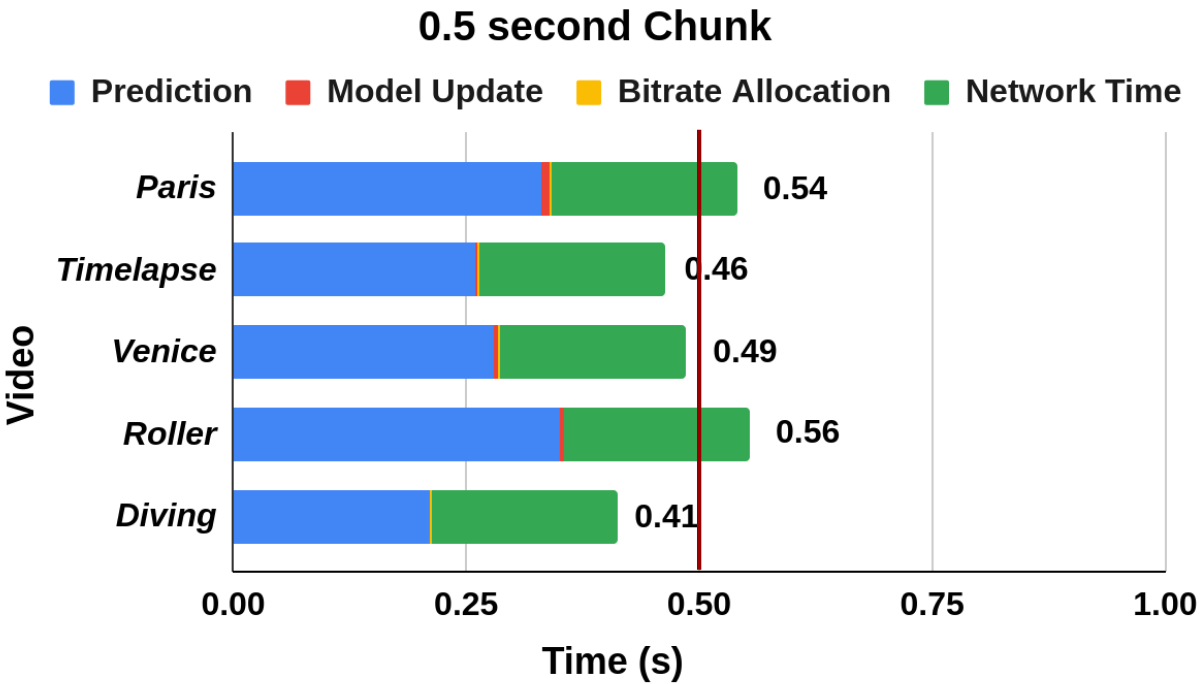}}
   \qquad
   \subfigure[]{\includegraphics[width =0.9\linewidth]{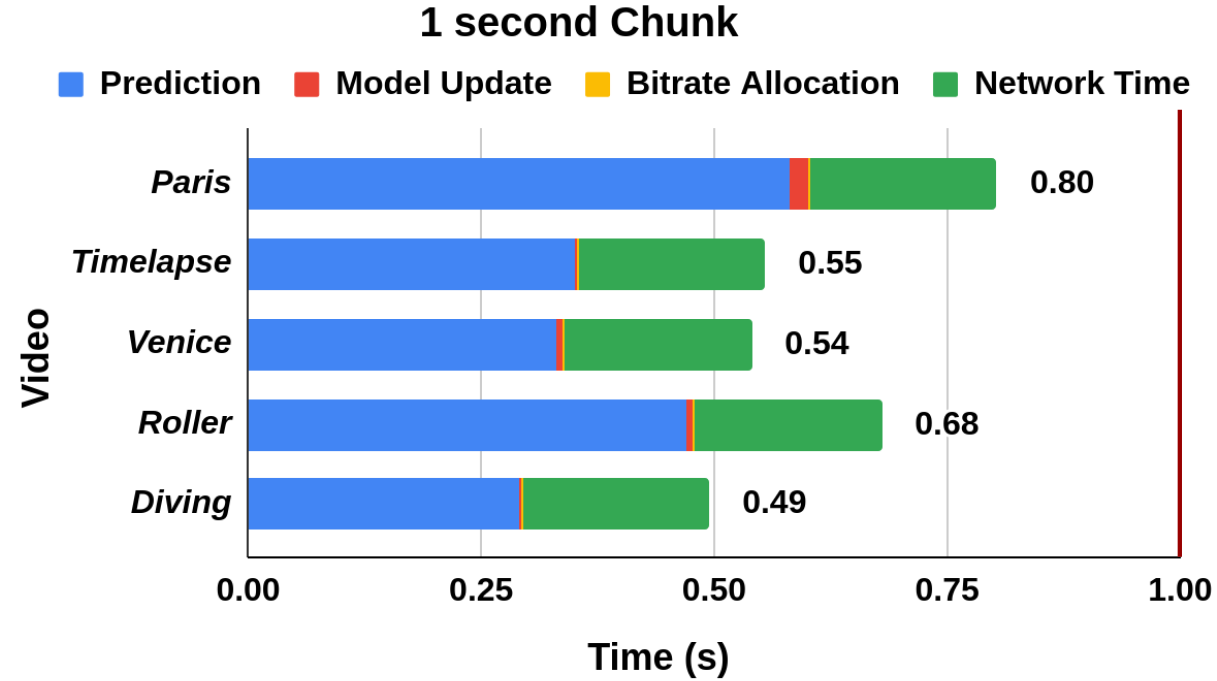}}
\vspace{-5mm}
\caption {\textit{Comparison of streaming times of chunks of duration 0.5 seconds and 1 second}}
\label{fig:fps_time}
\end{figure}
\par
However, although a chunk size of 0.5 seconds has a better QoE than a chunk of 1 second, it is not efficient from a video streaming point of view regarding the total streaming time of a chunk in a video. To argue over this fact, we find the average total streaming time for a chunk for the videos in ds1 averaged over all chunks for all users for chunk sizes of 0.5s and 1s (Figure \ref{fig:fps_time}). The streaming time is essentially composed of four parts: (1) model update time, (2) viewport prediction time, (3) bitrate allocation time, and (4) network time. The network time includes the time taken by the client to send a request and receive the video chunk from the server, decode the chunks, and stream them. As evident from the graphs, the model update and bitrate allocation time is meagre and not a bottleneck for any of the videos, whereas the prediction time is comparatively high. To facilitate our model even on 3G networks, we have considered the network latency to be 150 ms \cite{latency3G}. We use the video-streaming system developed to find the average decoding and streaming time of the chunks to be 50ms, making a total network time of 200ms for one chunk. We observe that for the chunk duration of 0.5s, the total streaming time is very close, and in some cases, exceeding 0.5 seconds. On the other hand, the total streaming time is comfortably under 1 second for a chunk duration of 1s, thus facilitating smooth video streaming. Also, MP4Box \cite{mpegdash, gpac} typically generates DASH segments of 1 second, also supported by Flare \cite{qian2018flare}, which further emphasizes on using 1 second over 0.5 seconds as our chunk size. Similar results were obtained on ds2. If a chunk of a certain time duration \textit{t} seconds takes more time than \textit{t} seconds to stream the chunk, it would essentially mean that the video will buffer at every chunk, which is not desired for a video-streaming system. Hence, our claim of using a chunk size of $fps$ (chunk duration 1 second) is justified.

\subsection{Enhancement over Individual Models}
We claimed that \textit{PARIMA} combines the benefits of the Passive-Aggressive Regression and ARIMA time series model. Thus, in order to justify our claim, we run viewport prediction for Passive-Aggressive Regression and ARIMA time series models separately and compare the results with \textit{PARIMA} model. 
\par
\begin{figure}[ht!]
        \centering
        \includegraphics[width=0.9\linewidth]{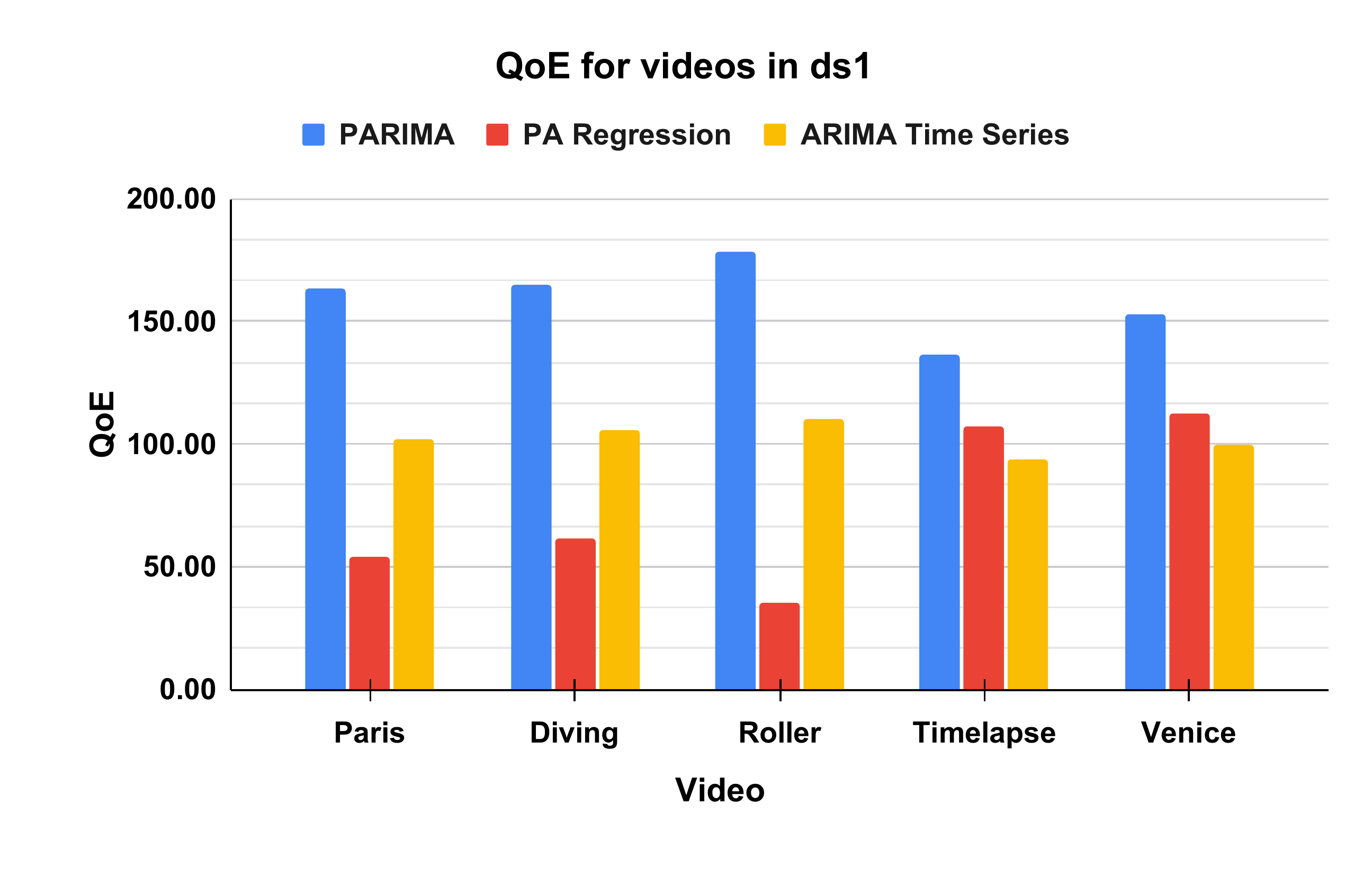}
        \caption{\textit{QoE of PARIMA, PA Regression and ARIMA Time Series Models}}
        \label{fig:pa_arima_parima}
\end{figure}

 For the Passive-Aggressive Regression model, the predicted viewport for a specific frame is computed using the objects' coordinates in that frame and the predicted viewport of the previous frame. For the ARIMA model, on the other hand, the predicted viewport for a chunk of frames is simply obtained using the actual viewports of the previous chunk, which is essentially the same as the `Intermediate Viewport' in Figure \ref{fig:parima}. We report QoE and Manhattan tile error for the three models for different videos averaged over all users. We plot the QoE for each model on dataset ds1 in Figure \ref{fig:pa_arima_parima} and tabulate the Manhattan Tile Error in Table \ref{tab:parima_pa_arima_table}. 

\begin{table}[ht!]
\centering
\begin{tabular}{|c|c|c|c|c|c|}
\hline
  \textbf{Model}  & \textbf{Paris} & \textbf{Diving} & \textbf{Roller} & \textbf{Timelapse} & \textbf{Venice}  \\ \hline
   \textbf{PARIMA} & 0.612 & 0.337 & 0.234 & 0.685 & 0.353  \\ \hline
   \textbf{PA} & 1.366 & 1.374 & 1.412 & 1.097 & 1.130 \\ \hline
   \textbf{ARIMA} & 0.643 & 0.368 & 0.305 & 0.779 & 0.438 \\ \hline
\end{tabular}
\caption{Average Manhattan Tile Error for PARIMA, PA Regression, ARIMA Time Series Models}
\label{tab:parima_pa_arima_table}
\end{table}

We can see that \textit{PARIMA} has a superior QoE and a lower tile error than the individual models. Passive-Aggressive Regression suffers from high tile errors due to the propagation of error over multiple frames. Since the predicted output of one frame is fed as an input to the next frame, the error in a certain viewport prediction gets propagated to the further frames. The Passive-Aggressive component shows great adaptivity due to its online nature and efficiently uses object trajectories to determine significance of various parts of the frames, leading to an overall useful model for viewport prediction. Similar inferences were obtained for ds2. ARIMA time series model, on the other hand, has low prediction errors and strengthens the Passive-Aggressive component when coupled into the \textit{PARIMA} model. The combined model has higher QoE and lower prediction errors than any of the two models, which validates the claim that the augmentation of the Passive Aggressive Regression and ARIMA models generates a superior model.  

\subsection{Measure of Object Contribution} \label{sec:obj_contri}
In our research, we claimed that a user's viewport depends upon the trajectory of the prime objects in the video. Since it also remains in the vicinity of the previous viewport, we use the last information viewport in the model effectively to predict the next chunk of viewports. To verify the above claim, we evaluate the percentage contribution of object trajectories in predicting the viewport. We present the results for the X-coordinate of the viewport since it typically has higher variability compared to the Y-coordinate. The object contribution essentially determines how significant they are in predicting the viewport of the user.



\begin{figure}[ht!]
 \centering
   \subfigure[\textbf{\footnotesize{\textit{Best in ds1: Venice}}}]{\includegraphics[width=0.45\linewidth]{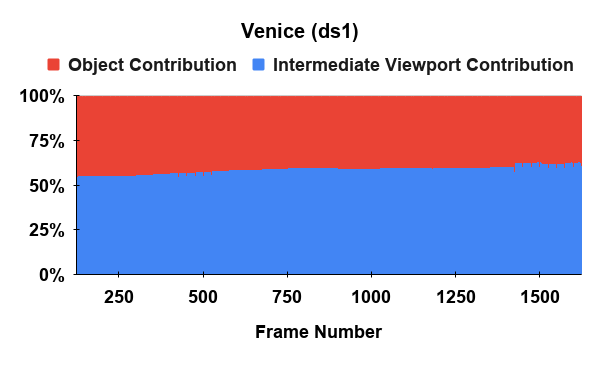}}
   \qquad
   \subfigure[\textbf{\footnotesize{\textit{Worst in ds1: Roller}}}]{\includegraphics[width =0.45\linewidth]{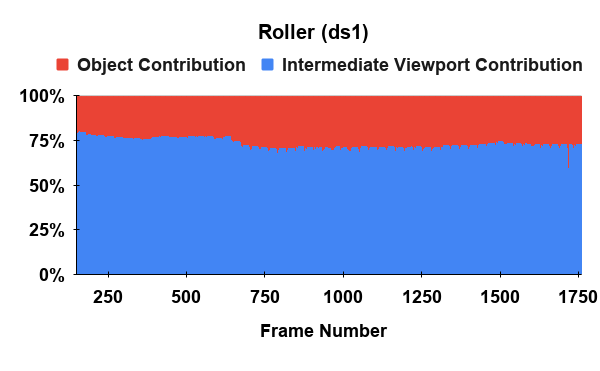}}
   \subfigure[\textbf{\footnotesize{\textit{Best in ds2: WeirdAI}}}]{\includegraphics[width=0.45\linewidth]{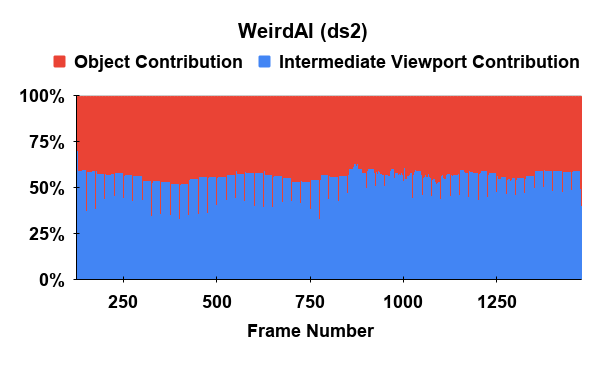}}
   \qquad
   \subfigure[\textbf{\footnotesize{\textit{Worst in ds2: Surfing}}}]{\includegraphics[width =0.45\linewidth]{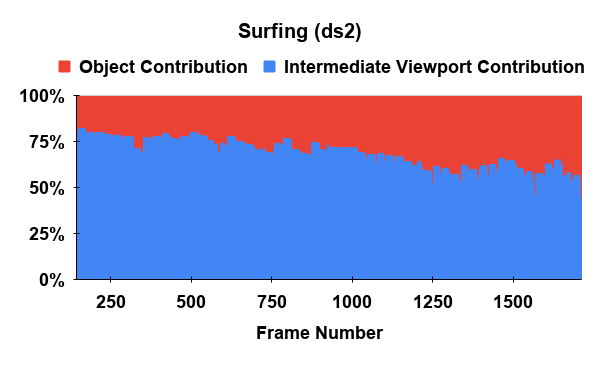}}
\vspace{-5mm}
\caption{Percentage contribution of objects and past viewport in final prediction from \textit{PARIMA}. Plots show the object contribution over the videos' length having the best and worst average object contribution \%. }
\label{fig:obj_contri}
\end{figure}

\begin{table}[ht!]
\begin{tabular}{|l|c|c|c|ll}
\cline{1-4}
 & \textbf{Video} & \textbf{\# Objects Detected} & \textbf{\begin{tabular}[c]{@{}c@{}}Avg. Object\\ Contribution (\%)\end{tabular}} &  &  \\ \cline{1-4}
\multicolumn{1}{|c|}{\multirow{5}{*}{\textbf{ds1}}} & \textbf{Paris} & 8 & 31.5 &  &  \\ \cline{2-4}
\multicolumn{1}{|c|}{} & \textbf{Diving} & 41 & 30.9 &  &  \\ \cline{2-4}
\multicolumn{1}{|c|}{} & \textbf{Roller} & 66 & 27.5 &  &  \\ \cline{2-4}
\multicolumn{1}{|c|}{} & \textbf{Timelapse} & 61 & 39.5 &  &  \\ \cline{2-4}
\multicolumn{1}{|c|}{} & \textbf{Venice} & 14 & 40.1 &  &  \\ \cline{1-4}
\multirow{9}{*}{\textbf{ds2}} & \textbf{Sandwich} & 41 & 27.4 &  &  \\ \cline{2-4}
 & \textbf{Skiing} & 77 & 31.5 &  &  \\ \cline{2-4}
 & \textbf{Alien} & 92 & 32.1 &  &  \\ \cline{2-4}
 & \textbf{WeirdAI} & 25 & 43.6 &  &  \\ \cline{2-4}
 & \textbf{Surfing} & 40 & 27.2 &  &  \\ \cline{2-4}
 & \textbf{War} & 397 & 34.5 &  &  \\ \cline{2-4}
 & \textbf{Cooking} & 1105 & 36.3 &  &  \\ \cline{2-4}
 & \textbf{Football} & 166 & 42.8 &  &  \\ \cline{2-4}
 & \textbf{Rhinos} & 77 & 27.9 &  &  \\ \cline{1-4}
 & \multicolumn{1}{l|}{\textbf{Average}} & \textbf{157} & \textbf{33.8} &  &  \\ \cline{1-4}
\end{tabular}
\caption{Average Percentage Object Trajectory Contribution}
\label{contri_tbl}
\end{table}

\begin{figure*}[ht!]
        \centering
        \includegraphics[width=0.9\linewidth]{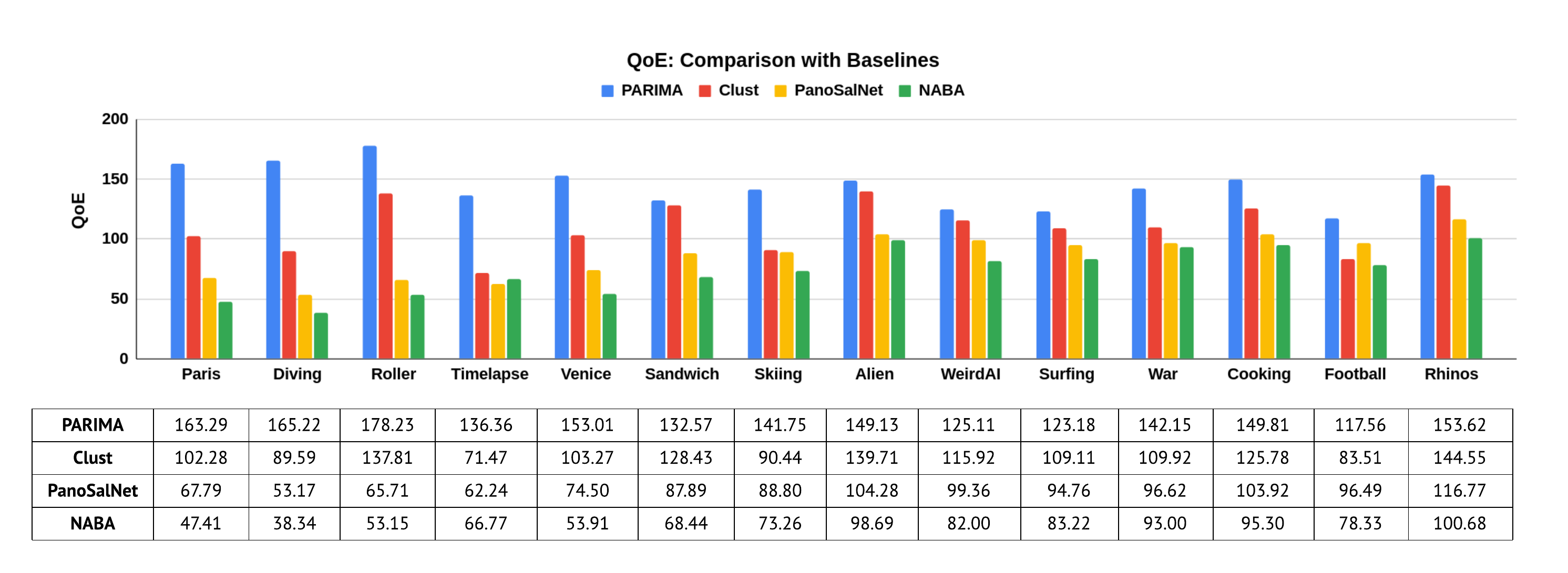}
        \caption{\textit{QoE comparison of PARIMA with PanoSalNet, Clust and NABA models}}
        \label{fig:qoe_baseline}
\end{figure*}

From Eq. \ref{eq:prediction}, we get the contribution of object trajectories in the prediction of the X coordinate of the viewport at frame $f$ as $\sum_{i=1}^{N_{obj}} \theta_{ix} . O_{Xif}$, while $\theta_{0x} + \theta_X . X_f^{ARIMA}$ is denoted as `Intermediate Viewport Contribution'. We find the proportion of contribution of object trajectory over the length of the video, averaged across all users which is shown in Figure \ref{fig:obj_contri} for the best and worst average object contribution for videos in ds1 and ds2. For all the videos in ds1 and ds2, we further report the average percentage contribution across all users for all frames in Table \ref{contri_tbl}.

Table \ref{contri_tbl} tabulates the number of objects detected by our object tracking algorithm and the average percentage contribution of object trajectories in predicting the user viewport. As observed, the percentage object contribution varies from 27.2\% to 43.6\% in our model for ds2 while ranging from 27.5\% to 40.1\% for ds1, with an average of 33.8\%. However, the standard deviation is only 5.58\%, and hence object trajectories contribute to roughly one-third of the viewport prediction on average. Even though \textit{Surfing} of ds2 shows the worst object contribution among all the other videos, it is not insignificant compared to the average of 33.8\%. From Figure \ref{fig:obj_contri}, we see that the object contribution of \textit{Surfing} increases as we play the video further, showing that the model learns that the viewport depends on the object trajectory. For \textit{WeirdAI}, which has the best average object contribution, a consistent value of around 40\% is maintained, similar to \textit{Venice}. 

\par Using the above results, it is evident that object trajectories play an important role in determining the viewport of the object, as they have significant contributions in the prediction. The use of object trajectories improving the PARIMA model over the ARIMA model in Table \ref{tab:parima_pa_arima_table} depicts that tile error reduces upon the inclusion of object trajectories, thus further justifying the claim that video content, in terms of object trajectories, play an essential role in viewport prediction. 

\subsection{Baseline Comparison}
Finally, to elicit the efficacy of our model, we evaluate \textit{PARIMA} against three state-of-the-art baselines: \textit{PanoSalNet}, \textit{Clust} and \textit{NABA}. We discussed in Section \ref{sec:testbed} how these baselines are congruent or effective in elucidating the effeciency of the PARIMA model.
\par \textit{PanoSalNet} and \textit{Clust} provide us with two viewport prediction models. We use them to predict viewports along with the bitrate allocation scheme discussed in Section \ref{bitrate_allocation_pyramid}. We find the Manhattan Tile Error and QoE for the videos in ds1 and ds2 averaged across all users across all chunks. The results for the Tile Error and QoE are shown in Table \ref{baseline_tbl} and Figure \ref{fig:qoe_baseline} respectively. 
\begin{table}[ht!]
\begin{tabular}{|l|c|c|c|c|}
\cline{1-5}
 & \textbf{Video} & \textbf{PARIMA} & \textbf{PanoSalNet} & \hspace{3mm}\textbf{Clust} \\ \cline{1-5}
\multicolumn{1}{|c|}{\multirow{5}{*}{\textbf{ds1}}} & \textbf{Paris} & 0.612 & 1.832 & \hspace{3mm}1.068  \\ \cline{2-5}
\multicolumn{1}{|c|}{} & \textbf{Diving} & 0.337 & 1.934 & \hspace{3mm}1.072  \\ \cline{2-5}
\multicolumn{1}{|c|}{} & \textbf{Roller} & 0.234 & 1.561 & \hspace{3mm}0.865  \\ \cline{2-5}
\multicolumn{1}{|c|}{} & \textbf{Timelapse} & 0.685 & 1.874 & \hspace{3mm}1.205  \\ \cline{2-5}
\multicolumn{1}{|c|}{} & \textbf{Venice} & 0.353 & 1.867 & \hspace{3mm}1.093  \\ \cline{1-5}
\multirow{9}{*}{\textbf{ds2}} & \textbf{Sandwich} & 0.144 & 1.645 & \hspace{3mm}1.097  \\ \cline{2-5}
 & \textbf{Skiing} & 0.156 & 2.691 & \hspace{3mm}1.337  \\ \cline{2-5}
 & \textbf{Alien} & 0.149 & 2.347 & \hspace{3mm}0.982  \\ \cline{2-5}
 & \textbf{WeirdAI} & 0.133 & 2.249 & \hspace{3mm}1.047  \\ \cline{2-5}
 & \textbf{Surfing} & 0.177 & 1.629 & \hspace{3mm}1.335  \\ \cline{2-5}
 & \textbf{War} & 0.178 & 1.661 & \hspace{3mm}1.089  \\ \cline{2-5}
 & \textbf{Cooking} & 0.159 & 1.355 & \hspace{3mm}1.484  \\ \cline{2-5}
 & \textbf{Football} & 0.175 & 2.511 & \hspace{3mm}1.130  \\ \cline{2-5}
 & \textbf{Rhinos} & 0.104 & 2.420 & \hspace{3mm}0.882  \\ \cline{1-5}
\end{tabular}
\vspace{3mm}
\caption{Tile Errors for PARIMA, PanoSalNet and Clust}
\label{baseline_tbl}
\end{table}

\par
As it can be observed from Figure \ref{fig:qoe_baseline}, \textit{PARIMA} exhibits a higher QoE than the two baselines for all the videos, with \textit{PanoSalNet} being the worse of the three. \textit{Clust} does not consider video contents while predicting viewport which, as discussed in Section \ref{sec:obj_contri}, play a significant role in \textit{PARIMA}. \textit{Clust} clusters the users based on viewport history and matches a new user to a cluster to predict the viewport. However, such a scheme can work for videos that particularly focuses on a character, as is the case for many videos in ds2. In ds1, however, the objects are spaced throughout the frame, and different users can have different viewing patterns. Hence, it becomes difficult to cluster users, or the new users can exhibit a different viewing pattern. It can be observed from Figure \ref{fig:qoe_baseline}, that for videos in ds1, \textit{Clust} performs much poorer than \textit{PARIMA} while the two methods are comparable for many videos in ds2. 

\par Since \textit{PanoSalNet} uses a saliency map to capture the video's content and employs LSTM based learning, its weights are not dynamically changed w.r.t. user preferences. Thus is often fails to categorize abrupt user behavior and allocates poor bitrate distribution among the tiles. This gets translated into a poorer prediction and hence a lower QoE. \textit{PARIMA} obtained an average improvement of 78.67\% in QoE over \textit{PanoSalNet} and 35.38\% over \textit{Clust}. In Table \ref{baseline_tbl}, we see similar results with \textit{PARIMA} performing better than the other two baselines.

\par \textit{Clust} can work well only when we have an existing head movement dataset of a sufficient set of users for efficient clustering, which would also essentially mean storing the head movements of all users. Similarly \textit{PanoSalNet} use existing head movement records to generate the saliency map. \textit{PARIMA}, on the other hand, just requires the object trajectory information, calculated one-time on the server side, and decouples the user viewport prediction from the viewports of other users, leading to lower memory/storage consumption, faster prediction and easy extensiblity to new videos. The model updates ensure that higher weights are given to the objects which the user is inclined to watch, leading to better user adaptation and hence, better QoE. 



\par
NABA model of bitrate allocation assumes no viewport prediction, and bitrate is allocated to each of the tiles in the chunk of frames equally. Hence, it is non-adaptive. As evident from Figure \ref{fig:qoe_baseline}, NABA exhibits the least QoE compared to the other models. \textit{PARIMA} performs comfortably better than NABA for all the videos with an average of 117.88\% improvement in adaptivity, hence verifying viewport-adaptivity for our model. 



\section{Conclusion and Future Work}
In this paper, we presented a novel tile-based viewport prediction model \textit{PARIMA}, that takes into account the video contents along with user head movement history to predict viewport for the future frames in the video. We have used object trajectories as a representative measure for the video content since they are exclusive to the video. We have shown through our experiments that while predicting the viewport, \textit{PARIMA} assigns around 34\% weightage to the object's position on an average. This verifies our claim that the viewport of a user depends not only on the previous viewport but also upon the trajectory of prime objects present in the video.
\par
Our system uses a 1-second chunk duration for viewport prediction and streaming. We have used a pyramid bitrate allocation scheme to allocate a higher bitrate to the predicted viewport and gradually decrease it as the tiles move away from the viewport. We evaluated \textit{PARIMA} and compared its performance against the current state-of-the-art video streaming solutions. Our evaluations show that \textit{PARIMA} offers better QoE relative to other state-of-the-art methods.
\par
In future, we plan to extend our work by predicting network inconsistencies and coupling the viewport-adaptive streaming methodology with network adaptivity. We also plan to incorporate audio channel as a supplemental representation of the video content, which includes various challenges including complex representation of the video and the presence of 3D audio. We plan to validate our work further by using more extensive datasets.

\bibliographystyle{ACM-Reference-Format}
\bibliography{main}


\begin{thebibliography}{42}


\ifx \showCODEN    \undefined \def \showCODEN     #1{\unskip}     \fi
\ifx \showDOI      \undefined \def \showDOI       #1{#1}\fi
\ifx \showISBNx    \undefined \def \showISBNx     #1{\unskip}     \fi
\ifx \showISBNxiii \undefined \def \showISBNxiii  #1{\unskip}     \fi
\ifx \showISSN     \undefined \def \showISSN      #1{\unskip}     \fi
\ifx \showLCCN     \undefined \def \showLCCN      #1{\unskip}     \fi
\ifx \shownote     \undefined \def \shownote      #1{#1}          \fi
\ifx \showarticletitle \undefined \def \showarticletitle #1{#1}   \fi
\ifx \showURL      \undefined \def \showURL       {\relax}        \fi
\providecommand\bibfield[2]{#2}
\providecommand\bibinfo[2]{#2}
\providecommand\natexlab[1]{#1}
\providecommand\showeprint[2][]{arXiv:#2}

\bibitem[\protect\citeauthoryear{??}{goo}{2017}]%
        {google-pixels}
 \bibinfo{year}{2017}\natexlab{}.
\newblock \showarticletitle{Bringing pixels front and center in VR video.
  \href{https://blog.google/products/google-ar-vr/bringing-pixels-front-and-center-vr-video/}{https://blog.google/products/google-ar-vr/bringing-pixels-front-and-center-vr-video/}}.
\newblock  (\bibinfo{year}{2017}).
\newblock


\bibitem[\protect\citeauthoryear{??}{fac}{2017}]%
        {facebook-vr}
 \bibinfo{year}{2017}\natexlab{}.
\newblock \showarticletitle{Facebook End-to-end optimizations for dynamic
  streaming.
  \href{https://engineering.fb.com/video-engineering/end-to-end-optimizations-for-dynamic-streaming/}{https://engineering.fb.com/video-engineering/end-to-end-optimizations-for-dynamic-streaming/}}.
\newblock  (\bibinfo{year}{2017}).
\newblock


\bibitem[\protect\citeauthoryear{??}{hev}{2018}]%
        {hevc}
 \bibinfo{year}{2018}\natexlab{}.
\newblock \showarticletitle{{HEVC(H.265)}: What is it and Why Should You Care?
  \href{https://blog.frame.io/2018/09/24/hevc-format-wars/}{https://blog.frame.io/2018/09/24/hevc-format-wars/}}.
\newblock  (\bibinfo{year}{2018}).
\newblock


\bibitem[\protect\citeauthoryear{??}{equ}{2020}]%
        {equirectangular}
 \bibinfo{year}{2020}\natexlab{}.
\newblock \showarticletitle{Equirectangular Projection.
  \href{https://en.wikipedia.org/wiki/Equirectangular_projection}{https://en.wikipedia.org/wiki/\\Equirectangular\_projection}}.
\newblock  (\bibinfo{year}{2020}).
\newblock


\bibitem[\protect\citeauthoryear{??}{pyq}{2020}]%
        {pyquaternion}
 \bibinfo{year}{2020}\natexlab{}.
\newblock \showarticletitle{Pyquaternion, Python module for representing and
  using quaternions.
  \href{http://kieranwynn.github.io/pyquaternion/}{http://kieranwynn.github.io/pyquaternion/}}.
\newblock  (\bibinfo{year}{2020}).
\newblock


\bibitem[\protect\citeauthoryear{??}{vrP}{2020}]%
        {vrProjector}
 \bibinfo{year}{2020}\natexlab{}.
\newblock \showarticletitle{vrProjector.
  \href{https://github.com/bhautikj/vrProjector}{https://github.com/bhautikj/vrProjector}}.
\newblock  (\bibinfo{year}{2020}).
\newblock


\bibitem[\protect\citeauthoryear{Adhikari and Agrawal}{Adhikari and
  Agrawal}{2013}]%
        {arima}
\bibfield{author}{\bibinfo{person}{Ratnadip Adhikari} {and}
  \bibinfo{person}{R.~K. Agrawal}.} \bibinfo{year}{2013}\natexlab{}.
\newblock \bibinfo{title}{An Introductory Study on Time Series Modeling and
  Forecasting}.
\newblock
\newblock
\showeprint[arxiv]{cs.LG/1302.6613}


\bibitem[\protect\citeauthoryear{Alface, Macq, and Verzijp}{Alface
  et~al\mbox{.}}{2012}]%
        {alface2012interactive}
\bibfield{author}{\bibinfo{person}{Patrice~Rondao Alface},
  \bibinfo{person}{Jean-Fran{\c{c}}ois Macq}, {and} \bibinfo{person}{Nico
  Verzijp}.} \bibinfo{year}{2012}\natexlab{}.
\newblock \showarticletitle{Interactive omnidirectional video delivery: A
  bandwidth-effective approach}.
\newblock \bibinfo{journal}{\emph{Bell Labs Technical Journal}}
  \bibinfo{volume}{16}, \bibinfo{number}{4} (\bibinfo{year}{2012}),
  \bibinfo{pages}{135--147}.
\newblock


\bibitem[\protect\citeauthoryear{Almquist, Almquist, Krishnamoorthi, Carlsson,
  and Eager}{Almquist et~al\mbox{.}}{2018}]%
        {almquist2018prefetch}
\bibfield{author}{\bibinfo{person}{Mathias Almquist}, \bibinfo{person}{Viktor
  Almquist}, \bibinfo{person}{Vengatanathan Krishnamoorthi},
  \bibinfo{person}{Niklas Carlsson}, {and} \bibinfo{person}{Derek Eager}.}
  \bibinfo{year}{2018}\natexlab{}.
\newblock \showarticletitle{The prefetch aggressiveness tradeoff in 360 video
  streaming}. In \bibinfo{booktitle}{\emph{Proceedings of the ACM MMSys 2018}}.
\newblock


\bibitem[\protect\citeauthoryear{Bao, Wu, Zhang, Ramli, and Liu}{Bao
  et~al\mbox{.}}{2016}]%
        {bao2016shooting}
\bibfield{author}{\bibinfo{person}{Yanan Bao}, \bibinfo{person}{Huasen Wu},
  \bibinfo{person}{Tianxiao Zhang}, \bibinfo{person}{Albara~Ah Ramli}, {and}
  \bibinfo{person}{Xin Liu}.} \bibinfo{year}{2016}\natexlab{}.
\newblock \showarticletitle{Shooting a moving target: Motion-prediction-based
  transmission for 360-degree videos}. In \bibinfo{booktitle}{\emph{2016 IEEE
  International Conference on Big Data (Big Data)}}. IEEE,
  \bibinfo{pages}{1161--1170}.
\newblock


\bibitem[\protect\citeauthoryear{{Bouzakaria}, {Concolato}, and {Le
  Feuvre}}{{Bouzakaria} et~al\mbox{.}}{2014}]%
        {mpegdash}
\bibfield{author}{\bibinfo{person}{N. {Bouzakaria}}, \bibinfo{person}{C.
  {Concolato}}, {and} \bibinfo{person}{J. {Le Feuvre}}.}
  \bibinfo{year}{2014}\natexlab{}.
\newblock \showarticletitle{Overhead and performance of low latency live
  streaming using MPEG-DASH}. In \bibinfo{booktitle}{\emph{Proceedings of the
  IISA 2014}}.
\newblock


\bibitem[\protect\citeauthoryear{Chen, Hu, Luo, Wang, and Wu}{Chen
  et~al\mbox{.}}{2020}]%
        {chen2020sr360}
\bibfield{author}{\bibinfo{person}{Jiawen Chen}, \bibinfo{person}{Miao Hu},
  \bibinfo{person}{Zhenxiao Luo}, \bibinfo{person}{Zelong Wang}, {and}
  \bibinfo{person}{Di Wu}.} \bibinfo{year}{2020}\natexlab{}.
\newblock \showarticletitle{SR360: boosting 360-degree video streaming with
  super-resolution}. In \bibinfo{booktitle}{\emph{Proceedings of the 30th ACM
  Workshop on Network and Operating Systems Support for Digital Audio and
  Video}}. \bibinfo{pages}{1--6}.
\newblock


\bibitem[\protect\citeauthoryear{Corbillon, De~Simone, and Simon}{Corbillon
  et~al\mbox{.}}{2017a}]%
        {corbillon2017360}
\bibfield{author}{\bibinfo{person}{Xavier Corbillon},
  \bibinfo{person}{Francesca De~Simone}, {and} \bibinfo{person}{Gwendal
  Simon}.} \bibinfo{year}{2017}\natexlab{a}.
\newblock \showarticletitle{360-degree video head movement dataset}. In
  \bibinfo{booktitle}{\emph{Proceedings of the ACM MMSys 2017}}.
\newblock


\bibitem[\protect\citeauthoryear{Corbillon, Simon, Devlic, and
  Chakareski}{Corbillon et~al\mbox{.}}{2017b}]%
        {corbillon2017viewport}
\bibfield{author}{\bibinfo{person}{Xavier Corbillon}, \bibinfo{person}{Gwendal
  Simon}, \bibinfo{person}{Alisa Devlic}, {and} \bibinfo{person}{Jacob
  Chakareski}.} \bibinfo{year}{2017}\natexlab{b}.
\newblock \showarticletitle{Viewport-adaptive navigable 360-degree video
  delivery}. In \bibinfo{booktitle}{\emph{Proceedings of the IEEE ICC 2017}}.
\newblock


\bibitem[\protect\citeauthoryear{Crammer, Dekel, Keshet, Shalev-Shwartz, and
  Singer}{Crammer et~al\mbox{.}}{2006}]%
        {crammer2006online}
\bibfield{author}{\bibinfo{person}{Koby Crammer}, \bibinfo{person}{Ofer Dekel},
  \bibinfo{person}{Joseph Keshet}, \bibinfo{person}{Shai Shalev-Shwartz}, {and}
  \bibinfo{person}{Yoram Singer}.} \bibinfo{year}{2006}\natexlab{}.
\newblock \showarticletitle{Online passive-aggressive algorithms}.
\newblock \bibinfo{journal}{\emph{Journal of Machine Learning Research}}
  \bibinfo{volume}{7}, \bibinfo{number}{Mar} (\bibinfo{year}{2006}),
  \bibinfo{pages}{551--585}.
\newblock


\bibitem[\protect\citeauthoryear{{Dasari}, {Bhattacharya}, {Vargas}, {Sahu},
  {Balasubramanian}, and {Das}}{{Dasari} et~al\mbox{.}}{2020}]%
        {parsec}
\bibfield{author}{\bibinfo{person}{M. {Dasari}}, \bibinfo{person}{A.
  {Bhattacharya}}, \bibinfo{person}{S. {Vargas}}, \bibinfo{person}{P. {Sahu}},
  \bibinfo{person}{A. {Balasubramanian}}, {and} \bibinfo{person}{S.~R. {Das}}.}
  \bibinfo{year}{2020}\natexlab{}.
\newblock \showarticletitle{Streaming 360-Degree Videos Using
  Super-Resolution}. In \bibinfo{booktitle}{\emph{Proceedings of the IEEE
  INFOCOM 2020}}.
\newblock


\bibitem[\protect\citeauthoryear{Dong, Ge, and Lee}{Dong et~al\mbox{.}}{2011}]%
        {latency3G}
\bibfield{author}{\bibinfo{person}{Wei Dong}, \bibinfo{person}{Zihui Ge}, {and}
  \bibinfo{person}{Seungjoon Lee}.} \bibinfo{year}{2011}\natexlab{}.
\newblock \showarticletitle{3G Meets the Internet: Understanding the
  Performance of Hierarchical Routing in 3G Networks}. In
  \bibinfo{booktitle}{\emph{Proceedings of the ITC 2011}}.
\newblock
\showISBNx{9780983628309}


\bibitem[\protect\citeauthoryear{Fan, Lee, Lo, Huang, Chen, and Hsu}{Fan
  et~al\mbox{.}}{2017}]%
        {fan2017fixation}
\bibfield{author}{\bibinfo{person}{Ching-Ling Fan}, \bibinfo{person}{Jean Lee},
  \bibinfo{person}{Wen-Chih Lo}, \bibinfo{person}{Chun-Ying Huang},
  \bibinfo{person}{Kuan-Ta Chen}, {and} \bibinfo{person}{Cheng-Hsin Hsu}.}
  \bibinfo{year}{2017}\natexlab{}.
\newblock \showarticletitle{Fixation prediction for 360 video streaming in
  head-mounted virtual reality}. In \bibinfo{booktitle}{\emph{Proceedings of
  the ACM NOSSDAV 2017}}.
\newblock


\bibitem[\protect\citeauthoryear{Fan, Lo, Pai, and Hsu}{Fan
  et~al\mbox{.}}{2019}]%
        {fan2019survey}
\bibfield{author}{\bibinfo{person}{Ching-Ling Fan}, \bibinfo{person}{Wen-Chih
  Lo}, \bibinfo{person}{Yu-Tung Pai}, {and} \bibinfo{person}{Cheng-Hsin Hsu}.}
  \bibinfo{year}{2019}\natexlab{}.
\newblock \showarticletitle{A Survey on 360° Video Streaming: Acquisition,
  Transmission, and Display}.
\newblock \bibinfo{journal}{\emph{ACM Computing Surveys (CSUR)}}
  \bibinfo{volume}{52}, \bibinfo{number}{4} (\bibinfo{year}{2019}),
  \bibinfo{pages}{71}.
\newblock


\bibitem[\protect\citeauthoryear{Gaddam, Riegler, Eg, Griwodz, and
  Halvorsen}{Gaddam et~al\mbox{.}}{2016}]%
        {gaddam2016tiling}
\bibfield{author}{\bibinfo{person}{Vamsidhar~Reddy Gaddam},
  \bibinfo{person}{Michael Riegler}, \bibinfo{person}{Ragnhild Eg},
  \bibinfo{person}{Carsten Griwodz}, {and} \bibinfo{person}{P{\aa}l
  Halvorsen}.} \bibinfo{year}{2016}\natexlab{}.
\newblock \showarticletitle{Tiling in interactive panoramic video: Approaches
  and evaluation}.
\newblock \bibinfo{journal}{\emph{IEEE Transactions on Multimedia}}
  \bibinfo{volume}{18}, \bibinfo{number}{9} (\bibinfo{year}{2016}),
  \bibinfo{pages}{1819--1831}.
\newblock


\bibitem[\protect\citeauthoryear{Guan, Zheng, Zhang, Guo, and Jiang}{Guan
  et~al\mbox{.}}{2019}]%
        {guan2019pano}
\bibfield{author}{\bibinfo{person}{Yu Guan}, \bibinfo{person}{Chengyuan Zheng},
  \bibinfo{person}{Xinggong Zhang}, \bibinfo{person}{Zongming Guo}, {and}
  \bibinfo{person}{Junchen Jiang}.} \bibinfo{year}{2019}\natexlab{}.
\newblock \showarticletitle{Pano: Optimizing 360 video streaming with a better
  understanding of quality perception}.
\newblock In \bibinfo{booktitle}{\emph{Proceedings of the ACM SIGCOMM 2019}}.
\newblock


\bibitem[\protect\citeauthoryear{Halford, Bolmier, Sourty, Vaysse, and
  Zouitine}{Halford et~al\mbox{.}}{2019}]%
        {creme}
\bibfield{author}{\bibinfo{person}{Max Halford}, \bibinfo{person}{Geoffrey
  Bolmier}, \bibinfo{person}{Raphael Sourty}, \bibinfo{person}{Robin Vaysse},
  {and} \bibinfo{person}{Adil Zouitine}.} \bibinfo{year}{2019}\natexlab{}.
\newblock \bibinfo{booktitle}{\emph{{creme}, a {P}ython library for online
  machine learning}}.
\newblock
\urldef\tempurl%
\url{https://github.com/MaxHalford/creme}
\showURL{%
\tempurl}


\bibitem[\protect\citeauthoryear{Han, Xu, Tao, and Gong}{Han
  et~al\mbox{.}}{2004}]%
        {han2004algorithm}
\bibfield{author}{\bibinfo{person}{Mei Han}, \bibinfo{person}{Wei Xu},
  \bibinfo{person}{Hai Tao}, {and} \bibinfo{person}{Yihong Gong}.}
  \bibinfo{year}{2004}\natexlab{}.
\newblock \showarticletitle{An algorithm for multiple object trajectory
  tracking}. In \bibinfo{booktitle}{\emph{Proceedings of the IEEE CVPR 2004}}.
\newblock


\bibitem[\protect\citeauthoryear{\href{http://faculty.smu.edu/tfomby/eco6375/BJ\%20Notes/ADF\%20Notes.pdf}{faculty.smu.edu}}{\href{http://faculty.smu.edu/tfomby/eco6375/BJ\%20Notes/ADF\%20Notes.pdf}{faculty.smu.edu}}{[n.d.]}]%
        {dickey-fuller}
\bibfield{author}{\bibinfo{person}{\href{http://faculty.smu.edu/tfomby/eco6375/BJ\%20Notes/ADF\%20Notes.pdf}{faculty.smu.edu}}.}
  \bibinfo{year}{[n.d.]}\natexlab{}.
\newblock \showarticletitle{Augmented Dickey-Fuller Unit Root Tests}.
\newblock  (\bibinfo{year}{[n.\,d.]}).
\newblock


\bibitem[\protect\citeauthoryear{Kua, Armitage, and Branch}{Kua
  et~al\mbox{.}}{2017}]%
        {kua2017survey}
\bibfield{author}{\bibinfo{person}{Jonathan Kua}, \bibinfo{person}{Grenville
  Armitage}, {and} \bibinfo{person}{Philip Branch}.}
  \bibinfo{year}{2017}\natexlab{}.
\newblock \showarticletitle{A survey of rate adaptation techniques for dynamic
  adaptive streaming over HTTP}.
\newblock \bibinfo{journal}{\emph{IEEE Communications Surveys \& Tutorials}}
  \bibinfo{volume}{19}, \bibinfo{number}{3} (\bibinfo{year}{2017}),
  \bibinfo{pages}{1842--1866}.
\newblock


\bibitem[\protect\citeauthoryear{Le~Feuvre, Concolato, and Moissinac}{Le~Feuvre
  et~al\mbox{.}}{2007}]%
        {gpac}
\bibfield{author}{\bibinfo{person}{Jean Le~Feuvre}, \bibinfo{person}{Cyril
  Concolato}, {and} \bibinfo{person}{Jean-Claude Moissinac}.}
  \bibinfo{year}{2007}\natexlab{}.
\newblock \showarticletitle{GPAC: Open Source Multimedia Framework}. In
  \bibinfo{booktitle}{\emph{Proceedings of the 15th ACM International
  Conference on Multimedia}} (Augsburg, Germany) \emph{(\bibinfo{series}{MM
  '07})}. \bibinfo{publisher}{Association for Computing Machinery},
  \bibinfo{address}{New York, NY, USA}, \bibinfo{pages}{1009–1012}.
\newblock
\showISBNx{9781595937025}
\urldef\tempurl%
\url{https://doi.org/10.1145/1291233.1291452}
\showDOI{\tempurl}


\bibitem[\protect\citeauthoryear{Mao, Netravali, and Alizadeh}{Mao
  et~al\mbox{.}}{2017}]%
        {mao2017neural}
\bibfield{author}{\bibinfo{person}{Hongzi Mao}, \bibinfo{person}{Ravi
  Netravali}, {and} \bibinfo{person}{Mohammad Alizadeh}.}
  \bibinfo{year}{2017}\natexlab{}.
\newblock \showarticletitle{Neural adaptive video streaming with pensieve}. In
  \bibinfo{booktitle}{\emph{Proceedings of the ACM SIGCOMM 2017}}.
\newblock


\bibitem[\protect\citeauthoryear{Nasrabadi, Samiei, and Prakash}{Nasrabadi
  et~al\mbox{.}}{2020}]%
        {clusterviewport}
\bibfield{author}{\bibinfo{person}{Afshin~Taghavi Nasrabadi},
  \bibinfo{person}{Aliehsan Samiei}, {and} \bibinfo{person}{Ravi Prakash}.}
  \bibinfo{year}{2020}\natexlab{}.
\newblock \showarticletitle{Viewport Prediction for 360° Videos: A Clustering
  Approach}. In \bibinfo{booktitle}{\emph{Proceedings of the ACM NOSSDAV
  2020}}.
\newblock
\showISBNx{9781450379458}


\bibitem[\protect\citeauthoryear{Nguyen and Yan}{Nguyen and Yan}{2019}]%
        {nguyen2019saliency}
\bibfield{author}{\bibinfo{person}{Anh Nguyen} {and} \bibinfo{person}{Zhisheng
  Yan}.} \bibinfo{year}{2019}\natexlab{}.
\newblock \showarticletitle{A saliency dataset for 360-degree videos}. In
  \bibinfo{booktitle}{\emph{Proceedings of the ACM MMSys 2019}}.
\newblock


\bibitem[\protect\citeauthoryear{Nguyen, Yan, and Nahrstedt}{Nguyen
  et~al\mbox{.}}{2018}]%
        {nguyen2018your}
\bibfield{author}{\bibinfo{person}{Anh Nguyen}, \bibinfo{person}{Zhisheng Yan},
  {and} \bibinfo{person}{Klara Nahrstedt}.} \bibinfo{year}{2018}\natexlab{}.
\newblock \showarticletitle{Your attention is unique: Detecting 360-degree
  video saliency in head-mounted display for head movement prediction}. In
  \bibinfo{booktitle}{\emph{Proceedings of the ACMMM 2018}}.
\newblock


\bibitem[\protect\citeauthoryear{Nguyen, Xu, Gao, Kankanhalli, Tian, and
  Yan}{Nguyen et~al\mbox{.}}{2013}]%
        {nguyen2013static}
\bibfield{author}{\bibinfo{person}{Tam~V Nguyen}, \bibinfo{person}{Mengdi Xu},
  \bibinfo{person}{Guangyu Gao}, \bibinfo{person}{Mohan Kankanhalli},
  \bibinfo{person}{Qi Tian}, {and} \bibinfo{person}{Shuicheng Yan}.}
  \bibinfo{year}{2013}\natexlab{}.
\newblock \showarticletitle{Static saliency vs. dynamic saliency: a comparative
  study}. In \bibinfo{booktitle}{\emph{Proceedings of the ACMMM 2013}}.
\newblock


\bibitem[\protect\citeauthoryear{Park, Bhattacharya, Yang, Dasari, Das, and
  Samaras}{Park et~al\mbox{.}}{2019}]%
        {park2019advancing}
\bibfield{author}{\bibinfo{person}{Sohee Park}, \bibinfo{person}{Arani
  Bhattacharya}, \bibinfo{person}{Zhibo Yang}, \bibinfo{person}{Mallesham
  Dasari}, \bibinfo{person}{Samir~R Das}, {and} \bibinfo{person}{Dimitris
  Samaras}.} \bibinfo{year}{2019}\natexlab{}.
\newblock \showarticletitle{Advancing User Quality of Experience in 360-degree
  Video Streaming}. In \bibinfo{booktitle}{\emph{Proceedings of the IFIP
  Networking 2019}}.
\newblock


\bibitem[\protect\citeauthoryear{Qian, Han, Xiao, and Gopalakrishnan}{Qian
  et~al\mbox{.}}{2018}]%
        {qian2018flare}
\bibfield{author}{\bibinfo{person}{Feng Qian}, \bibinfo{person}{Bo Han},
  \bibinfo{person}{Qingyang Xiao}, {and} \bibinfo{person}{Vijay
  Gopalakrishnan}.} \bibinfo{year}{2018}\natexlab{}.
\newblock \showarticletitle{Flare: Practical viewport-adaptive 360-degree video
  streaming for mobile devices}. In \bibinfo{booktitle}{\emph{Proceedings of
  the ACM MobiCom 2018}}.
\newblock


\bibitem[\protect\citeauthoryear{Redmon and Farhadi}{Redmon and
  Farhadi}{2018}]%
        {redmon2018yolov3}
\bibfield{author}{\bibinfo{person}{Joseph Redmon} {and} \bibinfo{person}{Ali
  Farhadi}.} \bibinfo{year}{2018}\natexlab{}.
\newblock \showarticletitle{Yolov3: An incremental improvement}.
\newblock \bibinfo{journal}{\emph{arXiv preprint arXiv:1804.02767}}
  (\bibinfo{year}{2018}).
\newblock


\bibitem[\protect\citeauthoryear{Stockhammer}{Stockhammer}{2011}]%
        {stockhammer2011dynamic}
\bibfield{author}{\bibinfo{person}{Thomas Stockhammer}.}
  \bibinfo{year}{2011}\natexlab{}.
\newblock \showarticletitle{Dynamic adaptive streaming over HTTP--: standards
  and design principles}. In \bibinfo{booktitle}{\emph{Proceedings of the ACM
  MMSys 2011}}.
\newblock


\bibitem[\protect\citeauthoryear{Sun, Mao, Zong, Liu, and Wang}{Sun
  et~al\mbox{.}}{2020}]%
        {sun2020flocking}
\bibfield{author}{\bibinfo{person}{Liyang Sun}, \bibinfo{person}{Yixiang Mao},
  \bibinfo{person}{Tongyu Zong}, \bibinfo{person}{Yong Liu}, {and}
  \bibinfo{person}{Yao Wang}.} \bibinfo{year}{2020}\natexlab{}.
\newblock \showarticletitle{Flocking-based live streaming of 360-degree video}.
  In \bibinfo{booktitle}{\emph{Proceedings of the 11th ACM Multimedia Systems
  Conference}}. \bibinfo{pages}{26--37}.
\newblock


\bibitem[\protect\citeauthoryear{Viitanen, Koivula, Lemmetti, Yl\"{a}-Outinen,
  Vanne, and H\"{a}m\"{a}l\"{a}inen}{Viitanen et~al\mbox{.}}{2016}]%
        {Kvazaar2016}
\bibfield{author}{\bibinfo{person}{Marko Viitanen}, \bibinfo{person}{Ari
  Koivula}, \bibinfo{person}{Ari Lemmetti}, \bibinfo{person}{Arttu
  Yl\"{a}-Outinen}, \bibinfo{person}{Jarno Vanne}, {and}
  \bibinfo{person}{Timo~D. H\"{a}m\"{a}l\"{a}inen}.}
  \bibinfo{year}{2016}\natexlab{}.
\newblock \showarticletitle{Kvazaar: Open-Source HEVC/H.265 Encoder}. In
  \bibinfo{booktitle}{\emph{Proceedings of the ACMMM 2016}}.
\newblock
\showISBNx{978-1-4503-3603-1}


\bibitem[\protect\citeauthoryear{Wu, Tan, Wang, and Yang}{Wu
  et~al\mbox{.}}{2017}]%
        {wu2017dataset}
\bibfield{author}{\bibinfo{person}{Chenglei Wu}, \bibinfo{person}{Zhihao Tan},
  \bibinfo{person}{Zhi Wang}, {and} \bibinfo{person}{Shiqiang Yang}.}
  \bibinfo{year}{2017}\natexlab{}.
\newblock \showarticletitle{A dataset for exploring user behaviors in VR
  spherical video streaming}. In \bibinfo{booktitle}{\emph{Proceedings of the
  ACM MMSys 2017}}.
\newblock


\bibitem[\protect\citeauthoryear{Xie, Zhang, and Guo}{Xie
  et~al\mbox{.}}{2018}]%
        {xie2018cls}
\bibfield{author}{\bibinfo{person}{Lan Xie}, \bibinfo{person}{Xinggong Zhang},
  {and} \bibinfo{person}{Zongming Guo}.} \bibinfo{year}{2018}\natexlab{}.
\newblock \showarticletitle{Cls: A cross-user learning based system for
  improving qoe in 360-degree video adaptive streaming}. In
  \bibinfo{booktitle}{\emph{Proceedins of the ACMMM 2018}}.
\newblock


\bibitem[\protect\citeauthoryear{Yilmaz, Javed, and Shah}{Yilmaz
  et~al\mbox{.}}{2006}]%
        {yilmaz2006object}
\bibfield{author}{\bibinfo{person}{Alper Yilmaz}, \bibinfo{person}{Omar Javed},
  {and} \bibinfo{person}{Mubarak Shah}.} \bibinfo{year}{2006}\natexlab{}.
\newblock \showarticletitle{Object tracking: A survey}.
\newblock \bibinfo{journal}{\emph{Acm computing surveys (CSUR)}}
  \bibinfo{volume}{38}, \bibinfo{number}{4} (\bibinfo{year}{2006}),
  \bibinfo{pages}{13--es}.
\newblock


\bibitem[\protect\citeauthoryear{Zare, Aminlou, Hannuksela, and Gabbouj}{Zare
  et~al\mbox{.}}{2016}]%
        {zare2016hevc}
\bibfield{author}{\bibinfo{person}{Alireza Zare}, \bibinfo{person}{Alireza
  Aminlou}, \bibinfo{person}{Miska~M Hannuksela}, {and} \bibinfo{person}{Moncef
  Gabbouj}.} \bibinfo{year}{2016}\natexlab{}.
\newblock \showarticletitle{HEVC-compliant tile-based streaming of panoramic
  video for virtual reality applications}. In
  \bibinfo{booktitle}{\emph{Proceedings of the ACMMM 2016}}.
\newblock


\bibitem[\protect\citeauthoryear{Zhang, Zhao, Bian, Liu, Song, and Li}{Zhang
  et~al\mbox{.}}{2019}]%
        {zhang2019drl360}
\bibfield{author}{\bibinfo{person}{Yuanxing Zhang}, \bibinfo{person}{Pengyu
  Zhao}, \bibinfo{person}{Kaigui Bian}, \bibinfo{person}{Yunxin Liu},
  \bibinfo{person}{Lingyang Song}, {and} \bibinfo{person}{Xiaoming Li}.}
  \bibinfo{year}{2019}\natexlab{}.
\newblock \showarticletitle{DRL360: 360-degree Video Streaming with Deep
  Reinforcement Learning}. In \bibinfo{booktitle}{\emph{Proceedings of the IEEE
  INFOCOM 2019}}.
\newblock


\end{thebibliography}

\end{document}